\documentclass{jpp}
\usepackage{graphicx}
\usepackage[T1]{fontenc}
\usepackage{amsmath}
\usepackage{bm}
\usepackage{natbib}
\usepackage{float}
\usepackage[colorlinks=true,
  linkcolor=black, citecolor=blue, urlcolor=blue]{hyperref}   % Modify colors of links
\usepackage[dvipsnames,svgnames,x11names,hyperref]{xcolor}  % Load colors for Tikz.
\usepackage[capitalize]{cleveref}

\usepackage{setspace}
\crefformat{subsection}{\S#1}

\shorttitle{GK turbulence using a moment based approach and advanced collision operators}
\shortauthor{A.C.D. Hoffmann, B.J. Frei, P. Ricci}

\title{Gyrokinetic simulations of plasma turbulence in a Z-pinch using a moment based approach and advanced collision operators}

\author{
A. C. D. Hoffmann \aff{1}\corresp{\email{antoine.hoffmann@epfl.ch}}, 
B. J. Frei \aff{1}, 
P. Ricci \aff{1}}
\affiliation{\aff{1} Ecole Polytechnique F\'ed\'erale de Lausanne (EPFL), Swiss Plasma Center (SPC), CH-1015 Lausanne, Switzerland
}
 \newcommand{\squareparenthesis}[1]{\left[#1\right]} 
 \newcommand{\roundparenthesis}[1]{\left(#1\right)} 
 \newcommand{\curlyparenthesis}[1]{\left\{#1\right\}} 
 \newcommand{\ex}{\bm{e}_x}
 \newcommand{\ey}{\bm{e}_y}
 \newcommand{\ez}{\bm{e}_z}

 \newcommand{\kperp}{k_\perp}
 \newcommand{\kpar}{k_\parallel}

 \newcommand{\vpar}{v_\parallel}
  
 \newcommand{\vperp}{v_\perp}
 \newcommand{\vtha}{v_{tha}}
 
 \newcommand{\FaM}{F_{aM}}

 \newcommand{\spara}{s_{\parallel a}}
 \newcommand{\upara}{u_{\parallel a}}

 \newcommand{\Pperpa}{P_{\perp a}}
 \newcommand{\Ppara}{P_{\parallel a}}

 \newcommand{\Rparpara}{R_{\parallel a}^\parallel}
 \newcommand{\Rperpperpa}{R_{\perp a}^{\perp}}
 \newcommand{\Rxa}{R_{x a}}
 \newcommand{\ddt}{\partial_t}

 \newcommand{\ddx}{\partial_x}
 \newcommand{\ddy}{\partial_y}
 
 \newcommand{\ddvpar}{\partial_{v\parallel}}

 \newcommand{\gradperp}{\nabla_\perp}

 \newcommand{\grad}{\nabla}
 
 \newcommand{\bvec}{\bm{b}}

\newcommand{\kernel}{\mathcal{K}}

\newcommand{\ExB}{\bm E\times\bm B}
\newcommand{\figref}[1]{Fig. \ref{#1}}
\newcommand{\modif}[1]{{#1}}
\newcommand{\modifii}[1]{{#1}}

\usepackage{natbib}
\usepackage{graphicx}
\usepackage{bm}
\usepackage{amsmath}
\numberwithin{equation}{section}
\usepackage{amssymb}
\usepackage{cancel}
\usepackage{float}
\usepackage{algorithmic}
\usepackage{algorithm}
\usepackage{multicol}

\begin{document}

\maketitle

\begin{abstract}
The first nonlinear gyrokinetic simulations obtained using a moment approach based on the Hermite-Laguerre decomposition of the distribution function are presented, implementing advanced models for the collision operator. 
Turbulence in a \modif{two-dimensional} Z-pinch is considered within a flux tube configuration. 
In the collisionless regime, our gyromoment approach shows very good agreement with nonlinear simulations carried out with the continuum gyrokinetic code GENE, even with fewer gyromoments than required for the convergence of the linear growth rate. 
By using advanced linear collision operators, the role of collisions in setting the level of turbulent transport is then analyzed.
The choice of collision operator model is shown to have a crucial impact when turbulence is quenched by the presence of zonal flows.
The convergence properties of the gyromoment approach improve when collisions are included.

\end{abstract}

\section{Introduction}
\modif{
% To efficiently study the dynamics the tokamak boundary, the region that encompasses the edge and scrape-off-layer (SOL), a gap must be bridged between fluid and gyrokinetic(GK) models.
% To study the dynamics of the tokamak boundary, including the edge and scrape-off-layer (SOL), a hybrid plasma model that blends the accuracy of the gyrokinetic (GK) model, while being computationally expensive, with the cost-effectiveness of a fluid approach,
% \modifii{Hybrid plasma models that combines the accuracy of the gyrokinetic (GK) model with the efficiency of a fluid approach at high collisionality must be found in order to study the dynamics of the tokamak boundary, the region that encompasses the edge and scrape-off-layer (SOL).}
\modifii{The understanding of the dynamics in the tokamak boundary, the region that encompasses the edge and scrape-off-layer (SOL), is crucial to predict the performance of future tokamak devices. 
While gyrokinetic (GK) models offer a proper model to simulate the plasma dynamics in core conditions, although at a high computational cost, fluid approaches are computationally less expensive but are limited to the high collisionality regime. 
% By blending the strength of both, a hybrid model could offer a cost-effective solution for studying the tokamak boundary dynamics.
}
To overcome the limitations of current models, \cite{Frei2020} propose an extension of the SOL drift-kinetic model presented in \cite{Jorge2017ACollisionality} and develop a GK model based on the projection of the velocity space dependence of the distribution function onto a Hermite-Laguerre polynomial basis.
Extending previous full-F gyrofluid models \citep{Strintzi2005NonlocalModel,Madsen2013Full-FModel,Held2020Pade-basedModels}, this set of fluid equations converges to the description of the evolution of the distribution function provided by the full-F GK Boltzmann equation, as the number of moments increases.
In \modifii{the Hermite-Laguerre} framework, advanced collision operators such as the full nonlinear Coulomb collision operator \citep{Jorge2019b}, as well as linearized ones \citep{Frei2021b} can be used to model collisional effects.}
\modif{In the $\delta f$ limit of the full-f model presented by \cite{Frei2020},
the Hermite-Laguerre decomposition can be interpreted as an extension of the previous gyrofluid model \citep{Brizard1992NonlinearPlasmas,Hammett1992FluidDynamics,Beer1995,Snyder2001AMicroturbulence,scott2005}} to an arbitrary number of moments.
This approach, also pursued by \cite{Mandell2018} with the GX code \citep{Mandell2018TheGx.readthedocs.io,Mandell2022GX:Design}, yields an infinite set of fluid equations for the basis coefficients, the gyromoments, which describe the deviations of the distribution function from a Maxwellian distribution.
The efficiency of the $\delta f$ \modifii{Hermite-Laguerre} gyromoment approach is demonstrated by \cite{Frei2022LocalMode} focusing on the linear properties of the ion temperature gradient (ITG) instability in the slab limit, as well as in a flux-tube geometry \citep{Frei2022Moment-BasedModel}, including the use of the linearized GK Landau form of the Fokker-Planck collision operator.
These works demonstrate the improvement of convergence properties of the gyromoment method with collisions, i.e. when deviations from a Maxwellian distribution function are reduced. 
In addition, even in the collisionless case, it is shown that the number of gyromoments needed for linear convergence is less than the number of grid points necessary for convergence \modif{in the state-of-the-art continuum GK code GENE \citep{Jenko2000}}.
\par
% \modif{To explore nonlinear regimes using a Hermite-Laguerre gyromoment approach for the first time,} 
\modif{Here, \modifii{nonlinear simulations are presented} for the first time using a Hermite-Laguerre gyromoment approach.}
We consider a local Z-pinch geometry which is characterized by a cylindrically symmetric plasma confined by a purely azimuthal, radially dependent, magnetic field with 
\modif{equilibrium radial gradients in temperature and density}.
In the presence of a background density gradient, an entropy mode \citep{Ricci2006} develops in the Z-pinch that can be modeled by using a local $\delta f$ GK approach with a kinetic treatment of the electrons.
\modif{This mode develops perpendicularly to the magnetic field and persists in the $k_\parallel=0$ limit, allowing simulations to be performed for a limited computational cost.}
\modif{While the Z-pinch geometry is considerably simpler than, e.g., the one in a tokamak (for instance, it does not have magnetic shear nor toroidal effects such as particle trapping), it still allows for the study of complex nonlinear phenomena, such as the emergence of zonal flows (ZF) \citep{Fujisawa2004, Diamond2005} that lead to the Dimits shift \citep{Dimits2000ComparisonsSimulations}, which role continues to challenge our understanding of tokamak physics.}
% Despite the simplicity of this configuration, turbulence in the Z-pinch geometry displays complex dynamics reminiscent of more complex magnetic topologies.
\par

The first nonlinear simulations in a Z-pinch, presented by \cite{Ricci2006a}, study the level of transport induced by the entropy mode as a function of the density gradient, showing that ZF can regulate the level of turbulent transport.
However, the effect of ZF can be \modif{reduced} either as the result of collisions, modeled in \cite{Ricci2006a} through a drift-kinetic (DK) Lorentz operator, or by a tertiary Kelvin-Helmholtz instability (KHI), destabilized in scenarios characterized by a sufficiently large density gradient drive.
These results are confirmed by \cite{Kobayashi2012} using a GK single-species collision operator described in \cite{Abel2008LinearizedTheory} and \cite{Barnes2009}.
\par
At low-density gradient drive, i.e. under the tertiary KHI instability threshold, transport regimes characterized by bursts rising from the competition between ZF collisional damping and quenching of the primary instability are identified and modeled with a predator-prey cycle by \cite{Kobayashi2015}.
In order to explore the mechanisms behind the ZF formation and damping, \cite{Ivanov2020ZonallyTurbulence} use a fluid-diffusive collision operator obtained by integration of the linearized Coulomb collision operator and derive a three-field, two-dimensional fluid model directly from the GK equation in a Z-pinch geometry, later extended to three dimensions \citep{Ivanov2022}.
This model includes first-order finite Larmor radius (FLR) effects in the long-wavelength, cold-ion limit and allows exploring the ZF dynamics within an analytical framework.
The simulations show good qualitative agreement with modified Hasegawa-Wakatani simulations \citep{Qi2020}.
Similarly, \cite{Hallenbert2021PredictingLimit} derive a fluid model in a Z-pinch geometry in the collisionless limit, including second-order FLR effects.
\modif{This allows the numerical prediction of the Dimits threshold, i.e. the gradient level below which transport is strongly reduced by the presence of ZF} \citep{Hallenbert2022}.
%Then, they predict numerically the Dimits transition, when the transport is strongly reduced by the presence of ZF when the density is below a density gradient threshold \citep{Hallenbert2022}.
The prediction is confirmed by comparison with GENE simulations.
% Finally it is worth noting that the former mentioned reduced models are treating electron dynamics with an adiabatic model which can restrain their field of applicability.
% In GK flux-tube simulations, it has been shown that adiabatic electron models may deliver erroneous Dimits threshold gradient value \citep{Mikkelsen2008} and transport level \citep{ Ball2020}.
%%%
\par
The present paper reports on the first nonlinear GK simulations carried out with the Hermite-Laguerre gyromoment approach using advanced collision operators \citep{gyacomo}. %\modif{\textsc{GYACOMO}, a Gyrokinetic Advanced Collision Moment solver \citep{gyacomo}}.
These simulations include nonlinear $\bm E \times \bm B$ advection, FLR effects of arbitrary order, kinetic electrons, and, leveraging the work in \cite{Frei2021b}, a set of advanced linear GK collision operators.
These operators include the single-species Dougherty model \citep{Dougherty1964}, the multi-species Sugama model \citep{Sugama2009LinearizedEquations}, the single-species pitch-angle scattering operator with a restoring momentum term, denoted as the Lorentz operator \citep{Helander2002CollisionalPlasmas}, and the Landau form of the multi-species Fokker-Planck model that we denote as Coulomb operator \citep{Rosenbluth1957,Hazeltine2003PlasmaConfinement}.
We consider a local $\delta f$ flux-tube approach that separates equilibrium and fluctuating quantities, assuming constant equilibrium gradients across the domain.
By imposing $k_\parallel=0$, we evolve the turbulent dynamics on a perpendicular plane.
\modif{This setup provides an ideal framework to compare the gyromoment model with a continuum code in a nonlinear turbulent regime, and to study the effect of advanced linearized collision models in ZF-dominated systems.}
\par
Our results demonstrate, first, the ability of the gyromoment approach to retrieve linear and nonlinear collisionless results obtained with the GK continuum code GENE.
In particular, we observe that the number of gyromoments needed for convergence increases while approaching the linear marginal stability conditions, and that underresolved collisionless simulations present predator-prey cycles, typically observed in collisional GK simulations \citep{Kobayashi2015} and fluid-reduced models \citep{Qi2020}.
The same dynamics is observed when increasing significantly the numerical dissipation acting on the velocity space in GENE.
Secondly, we present a set of simulations 
\modif{at different instability drives} in the collisionless limit and in the presence of collisions, which are modeled using the Dougherty, Sugama, Lorentz, and Coulomb collision operators.
The particle flux reveals a Dimits threshold in the collisionless limit.
For gradient levels above the Dimits threshold and at finite collisionality, we observe negligible differences between the different collision operators.
Shear flow stabilization effects are negligible and turbulence is fully developed.
The transport is well approximated by a mixing length argument, $\Gamma_x\sim \gamma^2/k^3$ \citep{Ricci2006a}, where $\Gamma_x$ is the saturated particle transport level along the radial direction, while $\gamma$ and $k$ are the peak linear growth rate and wavelength of the entropy mode, respectively.
Below the Dimits threshold,
%the mixing length argument is not sufficient to estimate and explain the nonlinear saturated transport level.
turbulence is quenched by ZF, which may be damped by collisions, and \modif{the choice of collision model affects significantly the transport level}.
% A ZF collisional damping study explains the differences observed between the collision operators.
\modif{A study of the ZF collisional damping provides and explanation for the differences observed between the collision operators.}
\par
The paper is organized as follows. In Sec. \ref{sec:model},  we briefly describe the nonlinear GK model in Z-pinch geometry and develop the gyromoment approach in this configuration.
Section \ref{sec:colless} presents linear and nonlinear benchmarks of the gyromoment approach with GENE in the collisionless limit.
The dependence of the transport level with the instability drive and the role of collisions is investigated in Sec. \ref{sec:collision}. The conclusions follow in Sec. \ref{sec:conclusion}.
\modif{In App. \ref{appendix:EHW}, we show that a gyrofluid model as well as an extended Hasegawa-Wakatani model can be obtained by properly truncating the gyromoment equation hierarchy.}

%\newpage
\section{Gyrokinetic model of a Z-pinch configuration based on the gyromoment model}
\label{sec:model}
%%%%%%%%%%%%%%% GYROKINETICS intro %%%%%%%%%%%%%%%%%%%%%%%%%%%%%%
In this section we present, first, the gyrokinetic (GK) model in the Z-pinch geometry, considering the local $\delta f$ flux-tube limit.
Second, we project the Z-pinch GK equation on a Hermite-Laguerre polynomial basis in velocity space, thus obtaining an infinite set of two-dimensional equations for the gyromoments, which we denote as the gyromoment equation hierarchy.
Finally, we present the numerical implementation of this hierarchy of equations.

\subsection{Gyrokinetic model in a Z-pinch configuration}
We consider the GK approach \citep{Catto1978LinearizedGyro-kinetics,Frieman1982NonlinearEquilibria,Hazeltine2003PlasmaConfinement} to study turbulence in a Z-pinch geometry.
Using the standard $\delta f$ approach, we decompose \modif{$f_a$}, the gyrocenter distribution function of species $a$ \modif{($a=e$ for electrons and $a=i$ for ions)}, as the sum of \modifii{a time-independent background} Maxwellian component and a perturbation, $f_a = \FaM + \delta f_a$, where the Maxwellian distribution for a species $a$ is defined as $\FaM = N_a/(\pi^{1/2}v_{tha})^{3} \exp (-m_a v_\parallel^2/2T_a-\mu B/T_a)$, with $\bm B=B \bvec$ the equilibrium magnetic field \modif{($B = |\bm B|, \bvec = \bm B/B$)}, $N_a$ the equilibrium density, $T_a$ the equilibrium temperature, $m_a$ the particle mass, \modif{$\bm v=\vpar \bvec + \bm \vperp$} the particle velocity, $\mu = m_a \vperp^2/B$ the particle magnetic moment, and $\vtha^2 = 2T_a/m_a$ the thermal velocity.
We assume small fluctuations, \modif{$\delta f_a/\FaM\sim \Delta \ll 1$, where the scaling parameter $\Delta$ measures the perturbation amplitude relative to the background \citep{Hazeltine2003PlasmaConfinement}}.
\par
\modif{We focus here on the Fourier representation of the perturbed gyrocenter distribution function at the gyrocenter position $\bm R$ and a time $t$,
\begin{equation}
g_a(\bm k,\vpar,\mu,t):=\int  \delta f_a(\bm R,\vpar,\mu,t) e^{-i\bm k \cdot \bm R} \mathrm d \bm R,    
\end{equation}
using Fourier modes $\bm k = \bm k_\perp + \kpar \bvec$.
The electrostatic GK Boltzmann equation determining the evolution of $g_a$ writes} \citep{Brizard2007}
\begin{equation}
   \ddt g_a  + i\omega_{Ba} g_a + \frac{1}{B}\{g_a+\FaM,J_0\phi\}= \sum_{b}C_{ab}.
    \label{eq:gyboeq}
\end{equation}
where we introduced the Poisson bracket operator, $\{f_1,f_2\}=\bvec \cdot (\nabla f_1 \times \nabla f_2)$ for two generic fields $f_1,f_2$, to describe the effect of the background density and temperature gradients, and of the quadratic nonlinearities\modif{, of order $\Delta^2$,} rising from the $\ExB$ drift.
In Eq. \eqref{eq:gyboeq} the magnetic drift frequency $i\omega_{Ba}$ contains the magnetic curvature and gradient drifts, i.e.
\begin{equation}
\omega_{Ba} =\bvec \times \frac{1}{\Omega_a}\squareparenthesis{\vpar^2 (\bvec\cdot\nabla)\bvec+\vperp^2 \grad B/B}\cdot \bm k_\perp.
\label{eq:iomegad}
\end{equation}
where $\Omega_a = q_a B/m_a$ is the cyclotron frequency with $q_a$ the particle charge.
Following previous work \citep{Ricci2006a,Ivanov2020ZonallyTurbulence}, we assume $\kpar = 0$ in Eq. \eqref{eq:gyboeq}, and therefore we consider a two-dimensional domain that extends perpendicularly to the magnetic field line.
The electrostatic potential $\phi$ is evaluated at the gyrocenter position through the gyroaveraging operator expressed, in Fourier space, with the zeroth order Bessel function of the first kind, $J_0=J_0(b_a)$ with $b_a =|\bm \kperp| |\bm \vperp|/\Omega_a$, containing FLR effects at all orders \modif{in $b_a$}.
Finally $C_{a,b}$ is the collision operator between species $a$ and $b$.
%
%POISSON'S EQUATION
%
\par
The electrostatic Poisson equation, in the quasi-neutrality limit, allows us to close the system by expressing the fluctuation of the electrostatic potential according to
\begin{equation}
    \sum_a \frac{q_a^2}{N_a T_a}\roundparenthesis{1- \Gamma_0(b_{a,th})}\phi = \sum_a q_a \int \mathrm d \bm v J_0 g_a,
    \label{eq:poissongk}
\end{equation}
where $b_{a,th} = (\kperp \vtha /\Omega_a)^2/2 $ and $\Gamma_0(x) = I_0(x) e^{-x}$ with $I_0$ the zeroth order modified Bessel function of the first kind.
\par

% \begin{figure}
%     \centering
%     \includegraphics[width = 0.5\linewidth]{figures/zpinch_illustration.eps}
%     \caption{Illustration of a magnetic field-line (blue) and a constant magnetic flux surface (gray mesh) of the considered Z-pinch magnetic equilibrium. The orientation of the magnetic gradient $\grad B$ and curvature $\bvec\cdot \grad \bvec$, the density gradient $\nabla N$ and the temperature gradient $\nabla T$ are also displayed. A snapshot of the electrostatic potential $\phi$ during the rise of the entropy mode instability for $\kappa_N=2.0$ and $\eta =0.25$ is presented on the periodic two-dimensional plane evolved by GYACOMO.}
%     \label{fig:zpinch_illustration}
% \end{figure}

\begin{figure}
    \centering
    \includegraphics[width = 0.5\linewidth]{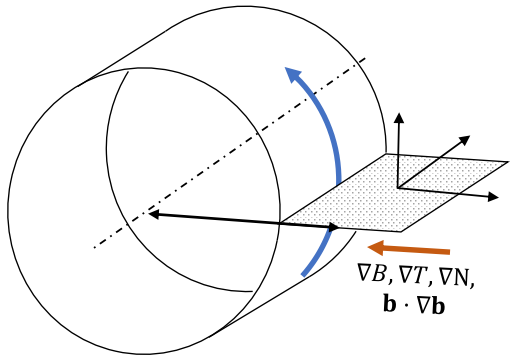}
    \vspace{-0.5cm}
    \caption{\modif{Illustration of the Z-pinch magnetic geometry considered here and the simulated perpendicular plane (gray area). 
    The field-aligned coordinate system and a magnetic field line $\bm B$ (blue arrow) are depicted. 
    We also indicate the direction of the density and temperature equilibrium gradients, $\nabla N$ and $\nabla T$, in addition to the magnetic equilibrium gradient and curvature, $\nabla B$ and $\bm b\cdot\nabla \bm b$, respectively (orange arrow). 
    The symmetry axis of the cylinder is represented by the dashed-dotted line and $L_B$ denotes the distance between the cylinder axis of symmetry and the flux-tube.}}
    \label{fig:zpinch_illustration}
\end{figure}

Equation \eqref{eq:gyboeq} is now simplified considering the Z-pinch magnetic field and geometry.
Using local field-aligned coordinates $(x,y,z)$, \modif{with $\ex$ the radial, $\ey$ the binormal and $\ez$ the azimuthal directions,} the Z-pinch magnetic field can be expressed as $\bm B = B \ez$.
The magnetic field presents a radial gradient, $\grad B/B = -1/L_B \ex$, and curvature, $(\bvec\cdot\grad)\bvec = -1/L_B \ex$, which are assumed constant within the flux-tube approach.
\modif{The length $L_B$ denotes the distance between the flux-tube and the symmetry axis of the Z-pinch (see Fig. \ref{fig:zpinch_illustration}).}
We also consider constant background density and temperature gradients, $\grad N_a/N_a = -1/L_N \ex$  and $\grad T_a / T_a = -1/L_T\ex$, for both electrons and ions.
\modifii{For comparison with common tokamak configuration, we note that, in the present geometry, all components of the metric tensor $g^{ij}$, for $i,j=x,y,z$, vanish except for $g^{xx}= 1$ and $g^{yy}=1/L_B^2$. 
In addition, one can express the Jacobian of the coordinate system as $J_{xyz}=L_B$ and the curvature operator as $[\bvec\times\grad B]\cdot\nabla = -B/L_B \ddy$ where $\ddy$ denotes the derivative in the $\ey$ direction}.
\modifii{We note that in the flux-tube framework, all background quantities ($B$,$N_a$ and $T_a$) and their associated gradients length ($L_B$, $L_N$ and $L_T$) are considered constant in time and in space.}
\par
\modif{
Throughout the rest of this work, we use the following dimensionless units.
The dimensionless parallel and perpendicular velocity coordinates are defined by $\spara = \vpar/\vtha$ and $x_a=\mu B /T_a$, respectively.
The perpendicular spatial scales are normalized to the sound Larmor radius $\rho_{s}=c_s/\Omega_i$, with $c_s=\sqrt{T_e/m_i}$ the sound speed.
Time is normalized to $L_B/c_s$.
The electrostatic potential is normalized to $T_e/e$ with $e$ the elementary charge, which allows us to define the normalized particle charge, $z_a=q_a/e$, as well.
We define the temperature and mass ratio $\tau_a=T_a/T_e$ and $\sigma_a=\sqrt{m_a/m_i}$, respectively, and we introduce the dimensionless density gradient drive, $\kappa_N=L_B/L_N$, the dimensionless temperature gradient drive, $\kappa_T=L_B/L_T$, and their ratio, $\eta = |\nabla\ln T|/|\nabla \ln N|$.
It is worth noting that the flux-tube limit in a Z-pinch is valid for $\rho_s/L_B\ll 1$.
}
\par

\modif{Considering purely perpendicular Fourier modes $\bm k = \bm \kperp = k_x \ex + k_y \ey$, the GK equations for $g_a$, Eq. \eqref{eq:gyboeq}, writes
\begin{align}
    \ddt g_a +\{g_a,J_0\phi\}+ \frac{\tau_a}{z_a}\squareparenthesis{\spara^2+\frac{1}{2}x_a} i k_y h_a +\squareparenthesis{\kappa_N +\kappa_T\roundparenthesis{\spara^2+x_a-\frac{3}{2}}}ik_y J_0 \phi = \sum_{b}C_{ab},
    \label{eq:gyboeq_ZP}
\end{align}
where we introduced the non-adiabatic part of the distribution perturbed function
\begin{equation}
    h_a(\bm k, \spara, x_a, t) = g_a(\bm k, \spara, x_a, t) + z_a/\tau_a J_0 \phi(\bm k, t).
\end{equation}
The Poisson bracket in the Z-pinch geometry writes as $\{f_1,f_2\}=\ddx f_1 \ddy f_2 - \ddy f_1 \ddx f_2$ in real space.
In the  Fourier, this yields a convolution expressed as
\begin{equation}
    % \{f_1,f_2\}=(ik_xf_1)*(ik_yf_2)-(ik_yf_1)*(ik_xf_2)
    \{f_1,f_2\}=\sum_{k_x',k_y'} k_x (k_y-k_y') f_1[\bm k-\bm k']f_2[\bm k] - k_y (k_x-k_x') f_1[\bm k-\bm k'] f_2[\bm k],
    \label{eq:poisson_bracket_fourier}
\end{equation}
}
Finally, we close our system with the dimensionless Poisson equation, i.e.
\begin{equation}
    \sum_a \frac{z_a^2}{\tau_a}\roundparenthesis{1- \Gamma_0(b_{a,th})}\phi = \sum_a z_a \int \mathrm d \spara \mathrm d x_a J_0 g_a.  
    \label{eq:GK_poisson}
\end{equation}
\par

%%%%%%%%%%%%%%% MOMENT HIERARCHY %%%%%%%%%%%%%%%%%%%%%%%%%%%%%%
\subsection{Nonlinear gyromoments hierarchy}
In order to solve Eq. \eqref{eq:gyboeq_ZP} by using the gyromoment framework, we expand the distribution function on a Hermite-Laguerre polynomial basis \citep{Jorge2017,Frei2020}, i.e.
\begin{equation}
    g_a(\bm k, \spara, x_a, t) = \sum_{p,j}N_a^{pj}(\bm k,t)H_p(\spara)L_j(x_a)\FaM(\spara,x_a).
    \label{eq:GM_expansion}
\end{equation}
\modif{In Eq. \eqref{eq:GM_expansion}, we introduce the \textit{gyromoment} of order $(p,j)$, i.e. the basis coefficient
\begin{equation}
N_a^{pj}(\bm k,t)=\int_0^\infty \mathrm d x_a \int_{-\infty}^\infty \mathrm d \spara g_a(\bm k, \spara, x_a, t) H_p(\spara) L_j(x_a),
\end{equation}
where }
\begin{equation}
    H_p(\spara)=\frac{(-1)^p }{\sqrt{2^p p!}} e^{\spara^2} \frac{ d^p}{ d\spara^p} e^{-\spara^2}
    \label{eq:hermite}
\end{equation}and
\begin{equation}
     L_j(x_a)=\frac{e^{x_a}}{j!}\frac{ d^j}{ dx_a^j}x_a^j e^{-x_a}
    \label{eq:laguerre}
\end{equation}
are the physicist's Hermite polynomial of order $p$ and the Laguerre polynomial of order $j$, respectively \citep{Gradshteyn2014TableProducts}.
The Hermite polynomials of Eq. \eqref{eq:hermite} are normalized such that $\int_{-\infty}^\infty \mathrm d \spara H_p H_{p'} e^{-\spara ^2}=\delta_{pp'}$ \modif{where $\delta_{pp'}$ denotes the Kronecker delta}.  
Similarly, the Laguerre polynomials satisfy the orthogonality relation $\int_{0}^\infty \mathrm d x_a L_j L_{j'} e^{-x_a}=\delta_{jj'}$.
%% Lines about history of Hermite polynomials in literature
\modifii{The use of Hermite polynomial projection is common in literature, particularly for projecting the one-dimensional velocity space Vlasov-Poisson system \citep{Armstrong1967NumericalEquation,Grant1967Fourier-HermiteLimit,Joyce1971NumericalEquation,Gibelli2006SpectralTerm,Parker2015Fourier-HermiteLimit}. 
On the other hand, Laguerre polynomials are not as frequently used in plasma physics compared to Hermite polynomials. 
Aside of spanning fluid equations \citep{Manas2017ImpactSimulations}, their main application is in expressing collision models in a spectral framework \citep{Brunner2000LinearTransport,Belli2012FullSimulations}.}
\par
We now project the Boltzmann GK equation, Eq. \eqref{eq:gyboeq_ZP}, onto the Hermite-Laguerre basis.
We expand the Bessel function of the first kind in terms of Laguerre polynomials as
\begin{equation}
    J_0 = J_0(\sqrt{l_a x_a}) = \sum_{n=0}^\infty \kernel_n(l_a)L_n(x_a),
\label{eq:bess_lag}
\end{equation}
with the kernel functions $\kernel_n(l_a)=l_a^n e^{-l_a}/n!$, being $l_a = \sigma_a^2\tau_a\kperp^2/2$ \citep{Frei2020}.
The projection of Eq. \eqref{eq:gyboeq_ZP} yields the gyromoment nonlinear hierarchy in a Z-pinch configuration, which can be expressed as
\begin{equation}
    \ddt N_a^{pj} + \mathcal{S}_a^{pj} + \mathcal{M}_a^{pj}  + \mathcal{D}_a^{pj} = \mathcal{C}_a^{pj},
    \label{eq:gmhierarchy}
\end{equation}
\modif{where the term related to the magnetic gradient and curvature drifts yields}
%$\mathcal{M}_a^{pj}=||\tau_a/z_a(\spara^2+x_a/2) i k_y g_a||_a^{pj}$ writes in terms of gyromoments
 \begin{align}
     \mathcal{M}_a^{pj} &= \frac{\tau_a}{z_a} i k_y \squareparenthesis{\sqrt{(p+1)(p+2)} n_a^{p+2,j} + (2p+1)n_a^{p,j} + \sqrt{p(p-1)}n_a^{p-2,j}}\nonumber\\
    &+ \frac{\tau_a}{z_a} i k_y \squareparenthesis{(2j+1)n_a^{pj} - (j+1)n_a^{p,j+1}-jn_a^{p,j-1}}.
    \label{eq:magnetic_drifts}
 \end{align}
% \modif{In Eq. \eqref{eq:magnetic_drifts}, we write explicitly the magnetic gradient
% and curvature strength dependencies, $\kappa_{Bg}$ and $\kappa_{Bc}$, respectively, which can be used for comparison with the extended Hasegawa-Wakatani model \citep{Hasegawa1983,Dewhurst2009TheTurbulence}, see Appendix \ref{appendix:EHW}.
% In this work we set as reference length the radius of the Z-pinch which implies $\kappa_{Bc} = \kappa_{Bg} = 1$.}
\modif{The term related to the density and temperature gradients writes}
%$\mathcal{D}_a^{pj}=||\kappa_N(1 +\eta(\spara^2+x_a-3/2) + (x_a+1)/\kappa_N)J_0 i k_y \phi||_a^{pj}$ yields
\begin{align}
    &\mathcal{D}_a^{pj}= -\kappa_N ik_y\phi[\kernel_j\delta_{p0}+\eta\kernel_j\frac{\sqrt{2}}{2}\delta_{p2} + \eta(2j\kernel_j-[j+1]\kernel_{j+1}-j\kernel_{j-1})\delta_{p0}].
    \label{eq:gradients_drifts}
\end{align}
\modif{In Eqs. \eqref{eq:magnetic_drifts} and \eqref{eq:gradients_drifts}, we introduce the non-adiabatic gyromoments
$n_a^{pj}(\bm k,t)=N_a^{pj}+z_a/\tau_a \kernel_j \phi \delta_{p0}$.}\\
The Hermite polynomial product rule, $\spara H_p = \sqrt{(p+1)/2}H_{p+1} + \sqrt{p/2}H_{p-1}$,
% \begin{equation}
% \spara H_p = \sqrt{(p+1)/2}H_{p+1} + \sqrt{p/2}H_{p-1},
% \end{equation}
and the Laguerre polynomial product rule, $x_a L_j = (2j+1)L_j-jL_j-(j+1)L_j,$
% \begin{equation}
% x_a L_j = (2j+1)L_j-jL_j-(j+1)L_j,
% \end{equation}
% as well as the derivative rule, $H_p'=\sqrt{2p}H_{p-1}$, 
are used to deduce Eqs. \eqref{eq:magnetic_drifts} and \eqref{eq:gradients_drifts}.
\par
%------------------ Nonlinear term
The nonlinear term related to the $\bm E\times \bm B$ drift is expressed in terms of gyromoments by using the Bessel-Laguerre decomposition, Eq. \eqref{eq:bess_lag}, and the Poisson bracket, Eq. \eqref{eq:poisson_bracket_fourier}, which yields
\begin{equation}
    % \mathcal{S}_a^{pj} =  -\sum_{n=0}^{\infty}\roundparenthesis{ik_x\kernel_n\phi}*\roundparenthesis{ik_y\sum_{s=0}^{n+j}d_{njs}N_a^{ps}} - \sum_{n=0}^{\infty}\roundparenthesis{ik_y\kernel_n\phi}*\roundparenthesis{ik_x\sum_{s=0}^{n+j}d_{njs}N_a^{ps}}.
    \mathcal{S}_a^{pj} =  \sum_{n=0}^{\infty}\curlyparenthesis{\sum_{s=0}^{n+j}d_{njs} N_a^{ps},\kernel_n\phi}.
   \label{eq:sapj_def}
\end{equation}
To obtain Eq. \eqref{eq:sapj_def}, we expressed the product of two Laguerre polynomials as a sum of single polynomials using the identity
\begin{equation}
L_jL_n=\sum_{s=0}^{n+j}d_{njs}L_s
\label{eq:lagprod}
\end{equation}
with
\begin{equation}
    d_{njs} = \sum_{n_1=0}^n\sum_{j_1=0}^j\sum_{s_1=0}^s \frac{(-1)^{n_1+j_1+s_1}}{n_1!j_1!s_1!}\binom{n}{n_1}\binom{j}{j_1}\binom{s}{s_1}.
    \label{eq:dnjs}
\end{equation}
\modif{This choice differs from the representation in the GX code that evaluates the Laguerre product with a pseudo-spectral algorithm in the velocity space \citep{Mandell2018, Mandell2022GX:Design}.}
\par
%%%%%%%%%%%%%%% POISSON  %%%%%%%%%%%%%%%%%%%%%%%%%%%%%%
The Poisson equation, Eq. \eqref{eq:GK_poisson}, is also projected onto the Hermite-Laguerre basis.
This yields \citep{Frei2020}
\begin{equation}
    \squareparenthesis{\sum_a \frac{z_a^2}{\tau_a}\roundparenthesis{1-\sum_{n=0}^{\infty}\kernel^2_n}}\phi = \sum_a z_a \sum_{n=0}^{\infty}\kernel_n N_a^{0n},
    \label{eq:poisson_moments}
\end{equation}
where the quasi-neutrality approximation is used, i.e. $(k_\perp\lambda_D)^2\ll 1$ with $\lambda_D$ the Debye length.
\modif{In the collisionless limit ($\mathcal C_a^{pj}=0$), the gyromoment hierarchy, Eq. \eqref{eq:gmhierarchy}, combined with the Poisson equation, Eq. \eqref{eq:poisson_moments}, can be considered as an extension of the gyrofluid model to an arbitrary number of moments.
In App. \ref{appendix:EHW}, we demonstrate that the formerly derived gyrofluid model in \cite{Brizard1992NonlinearPlasmas} can be retrieved by properly truncating the collisionless gyromoment hierarchy. 
We also show how an extended Hasegawa-Mima model \citep{Hasegawa1978Pseudo-three-dimensionalPlasma,Dewhurst2009TheTurbulence} can be obtained.
}
\par
%%%%%%%%%%%%%%%  TRANSPORT
Finally, we note that we characterize the turbulent transport in a Z-pinch by considering the dimensionless ion particle flux, $\bm \Gamma = n_i \bm v_{E\times B}$, with $\bm v_{E\times B}=-\nabla \phi \times \bvec$ the $\ExB$ velocity and $ n_i=\sum_{n=0}^\infty \kernel_n N_i^{0n}$ the ion particle density perturbation.
In the following, we analyze the time series of the spatially averaged radial ion particle flux, $\Gamma_x(t) = \langle \bm \Gamma \cdot \ex \rangle_{xy}$, which can be expressed, using the Fourier modes of the gyromoments, as
\begin{equation}
    \Gamma_x(t) =\sum_{k_x,k_y} (ik_y\phi)^* \sum_{n=0}^{\infty}\kernel_n N_i^{0n}.
    \label{eq:particle_flux}
\end{equation}
\modif{The saturated radial particle transport, $\Gamma_x^{\infty}$, is analyzed by evaluating the convergence of the quantity $\bar{\Gamma_x}(t) = \int_{t_0}^{t}\Gamma_x(t')dt'/(t-t_0)$ as $t$ increases, considering $t_0$ sufficiently large that the initial transient present in the simulation is not considered. 
The value of $\bar \Gamma_x(t)$ provides an estimate for the saturated transport level, $\Gamma_x^{\infty} = \underset{t\rightarrow \infty}{\lim}\bar\Gamma_x(t)$.}

%------------------ Collision
\subsection{Linear collision operators}
\modif{
The $\mathcal C_a^{pj}$ term in Eq. \eqref{eq:gmhierarchy} represents the effect of collisions through the projection of a collision operator model onto the Hermite-Laguerre basis.
Any linearized Fokker-Planck collision operator can be written as the sum of a test part $C^T_{ab}$ and a field part $C^F_{ab}$, i.e. $C_{ab} = C^T_{ab} +C^F_{ab}$ \citep{Helander2002CollisionalPlasmas,Hazeltine2003PlasmaConfinement},
with $C^T_{ab} = C(f_a,F_{bM})$ and $C^F_{ab} = C(\FaM,f_b)$ for any species $a$ and $b$. 
\par
We consider here the Coulomb, Sugama, Lorentz, and Dougherty operators.
For the case of the Coulomb collision operator, we introduce the Rosenbluth potentials, $H(f)=2\int d^3 v' f(\bm v')/|\bm v-\bm v'|$ and $G(f)=\int d^3 v' |\bm v - \bm v'| f(\bm v')$, as well as the phase space coordinates $(\bm r, v, \xi,\theta)$, where $\bm r$ denotes the particle position, $v$ the magnitude of its velocity, $\xi=v_\parallel/v$ the pitch-angle and $\theta$ the gyro-angle.
In this framework, the test part of the Coulomb collision operator can be expressed as \citep{Frei2021b}
\begin{align}
    C_{ab}^T =& \frac{m_a\nu_{ab}}{N_b}\left\{ \squareparenthesis{\frac{2}{v^2}G(F_{bM} )
    + \roundparenthesis{1 - \frac{m_a}{m_b}\partial_v H(F_{bM})}} \partial_v f_a \right.
    \nonumber\\&
    \left. - \frac{1}{v^3} \partial_v G(F_{bM}) \mathcal{L}^2 f_a
    + \partial^2_v G(F_{bM})\partial_v^2 f_a + \frac{m_a}{m_b}8\pi F_{bM} f_a \right\},
    \label{eq:FC_test}
\end{align}
where $\mathcal{L}f=\partial_\xi[(1-\xi^2)\partial_\xi f] + \partial^2_\theta f/(1-\xi^2)$ is the pitch-angle operator.
On the other hand, the field part yields
\begin{equation}
    C_{ab}^F=\frac{2\nu_{ab}\vtha \FaM}{N_b} \squareparenthesis{2\frac{v^2}{\vtha^2}\partial_v^2 G(f_b) - H(f_b) - \roundparenthesis{1-\frac{m_a}{m_b}}v\partial_v H(f_b) + \frac{m_a}{m_b}4\pi \vtha^2 f_b}.
    \label{eq:FC_field}
\end{equation}
It is worth noting that the computation of the field term is particularly costly because of the velocity integrals of the perturbed distribution function contained in the Rosenbluth potentials.
However, the projection of this operator on the Hermite-Laguerre basis enables the expression of these integrals as a linear combination of gyromoments.
In this work, the Coulomb operator refers to the gyro-averaged version of the linearized Fokker-Planck collision operator, Eqs. \eqref{eq:FC_test} and \eqref{eq:FC_field}.

% Sugama
The Sugama collision model \citep{Sugama2009LinearizedEquations} is a multi-species generalization of the Abel operator \citep{Abel2008LinearizedTheory}.
While the Sugama operator considers the test part of the Fokker-Planck operator, Eq. \eqref{eq:FC_test}, which includes pitch-angle scattering and energy diffusion, the field term is replaced by an ad-hoc term derived from a fluid approach to conserve particle, momentum, energy and satisfy the H-theorem.

% Pitch-angle scattering
The Lorentz model considers like-particle collisions, i.e. $a=b$, and it is based on the small mass ratio limit, which simplifies the test part of Eq. \eqref{eq:FC_test} to the pitch-angle operator term only.
The field part is adapted to conserve particles, momentum, and energy.
\cite{Ricci2006a} observe that this operator does not provide sufficient damping to avoid the use of artificial dissipation in nonlinear simulations.
This is in contrast to the Sugama and Abel operators.

% Dougherty
Finally, the Dougherty model consists of kinetic and spatial second-order diffusion terms \citep{Lenard1958} with corrections involving the density, velocity, and temperature fluid moments in order to conserve particle, momentum, and energy.
}
The details of the Dougherty, Sugama, Lorentz, and Coulomb GK operators as well as their projection onto the Hermite-Laguerre basis can be found in \cite{Frei2021b}.
We set the intensity of the collisions through the normalized ion-ion collision frequency $\nu$.
The collision frequencies among the different species are thus given by $\nu_{ii}=\nu$, $\nu_{ee}=\sigma_e \tau_e^{3/2}\nu$, $\nu_{ei}=\nu$ and $\nu_{ie}=\sigma_e \tau_e^{3/2}\nu$.
\par

\subsection{Numerical approach}
%%%%%%%%%%%%%%% IMPLEMENTATION start %%%%%%%%%%%%%%%%%%%%%%%%%%%%%%
To solve Eq. \eqref{eq:gmhierarchy} numerically, we evolve a finite set of gyromoments $N_a^{pj}(\bm k,t)$ with $0\leq p \leq P$ and $0\leq j \leq J$ and consider the Fourier modes with $k_x=m\Delta k_x$, with $0\leq m\leq M$, and $k_y=n\Delta k_y$, with $-N/2+1\leq n \leq N/2$, using a standard explicit fourth-order Runge-Kutta time-stepping scheme.
In the Z-pinch geometry, the gyromoments hierarchy decouples odd and even Hermite gyromoments, which is a consequence of the $k_\parallel=0$ assumption.
This allows us to evolve only the even gyromoments $N_a^{pj}$, with $p= {2l,l\in\mathbb{N}}$.
The hierarchy is closed by using a simple truncation, i.e. $N_a^{p,j}=0$ for all $p>P$ or $j>J$.
\modif{In the following, we denote this truncated gyromoment set as a $(P,J)$ basis.}
The use and analysis of more advanced closure schemes, e.g. the semi-collisional closure proposed by \cite{Zocco2011} and \cite{Loureiro2016Viriato:Dynamics}, are left for future work.
\par
Focusing on the nonlinear term in Eq. \eqref{eq:sapj_def}, we first observe that any truncation of the sum over $s$ must be avoided in order to prevent polynomial aliasing.
Hence, to guarantee the exact Laguerre product identity in Eq. \eqref{eq:lagprod}, we truncate the sum over $n$ in Eq. \eqref{eq:sapj_def} to $n\leq J-j$.
Second, we note that the computation of the $d_{njs}$ coefficients is challenging since they involve sums and differences of large numbers.
To avoid the overflow of the floating point representation, we use an arbitrary precision library for our calculations \citep{Smith1991AlgorithmArithmetic}.
Finally, we note that the convolutions in Fourier space are treated with a conventional pseudo-spectral method, i.e. the backward fast Fourier transform \citep{FFTW05} of the fields to convolve, the multiplication in real space and the forward fast Fourier transform of the result, including the usual 2/3 Orszag rule for anti-aliasing \citep{Orszag1971}.
\par
Regarding the collision operators, the Dougherty operator, which has a light computational cost, is directly implemented in the gyromoment hierarchy.
On the other hand, the evaluation of the Sugama, Lorentz pitch-angle, and Coulomb collision terms is reduced to a four-dimensional matrix-vector operation, i.e. the $pj$-th collision term is written as $C_{a}^{pj}=\sum_b\sum_{p'=0}^{P_b}\sum_{j'=0}^{J_b}\mathcal{C}_{ab}^{pj,p'j'} N_{b}^{p'j'}$ with a precomputed collision matrix $\mathcal{C}_{ab}^{pj,p'j'}$ of size $(P_a\times J_a)\times(P_{b}\times J_{b})$.
The projection of the collision operators on the Hermite-Laguerre basis and the details of the computation of the matrix coefficients for each collision operator considered in the present work can be found in \cite{Frei2021b}.
It is worth noting that the GK corrections create a $\kperp$ dependence of the matrix coefficients, i.e. $\mathcal{C}_{ab}^{pj,p'j'}=\mathcal{C}_{ab}^{pj,p'j'}(\kperp)$, that calls for the precomputation of the coefficients for each $\kperp$ present in the simulations.
\modif{The computational cost of evaluating the GK matrix coefficients increases with $\kperp$.
Indeed, at large $\kperp$, accurate FLR effects ask for larger bounds in the truncated sums used to approximate Bessel functions and basis transformations (see \cite{Frei2021b} for more details).
We ensured the convergence of our matrix evaluation by analyzing the eigenvalue spectrum and the matrix symmetry.}
%%%%%%%%%%%%%%% IMPLEMENTATION end %%%%%%%%%%%%%%%%%%%%%%%%%%%%%%

% \newpage
\section{Collisionless limit and comparison with the GENE code}
\label{sec:colless}
In the present section, we analyze the results of the gyromoment simulations in the collisionless limit and demonstrate the ability of this approach to retrieve the results of the continuum gyrokinetic code GENE in a collisionless two-dimensional Z-pinch configuration considering an equilibrium plasma with $\tau=1$, and a realistic \modif{electron-proton} mass ratio, $\sigma= \sqrt{m_e/m_i}= 0.023$.
\modif{GENE simulations are set up by following closely \cite{Hallenbert2022}.}
 \par
 \subsection{Entropy mode instability}
%%%%LINEAR
At density gradients below the magneto-hydrodynamic (MHD) interchange instability threshold, a small-scale non-MHD instability, the entropy mode, can be destabilized in the Z-pinch configuration.
The region of stability of the entropy mode is presented in \cite{Ricci2006}.
Our analysis focuses on density gradient values $1.6\leq \kappa_N \leq2.5$, while the temperature density gradient ratio is constant, $\eta =0.25$.
This parameter encompasses an unstable region of the entropy mode, which extends, indeed, from $\kappa_N\simeq 2.5$, where the ideal MHD interchange mode is destabilized, to $\kappa_N\simeq 1.6$, \modif{which is close the analytical stability limit found by \cite{Ricci2006}, that is $\kappa_N=\pi/2$ for $\eta = 0$}.
\par
We start the analysis of the collisionless case by focusing on the linear growth rate of the entropy mode.
\modif{The entropy mode instability growth rate is obtained by solving the initial value problem associated with the gyromoment hierarchy, Eq. \eqref{eq:gmhierarchy}, coupled to the Poisson equation, Eq. \eqref{eq:poisson_moments}, and where the nonlinear terms, developed in Eq. \eqref{eq:sapj_def}, are neglected.
In particular, we evolve $\phi$ and $N_a^{pj}$ as a function of $k_y$ modes, setting $k_x=0$ where the entropy mode growth rate peaks \citep{Ricci2006}.
We compute the growth rates, $\gamma(k_y)$, by fitting the slope of the time evolution of $\ln |\phi_{k_y}|$ over a time window.
The convergence is tested by checking that the results are independent of the size of the time window.}\par
\begin{figure}
    \centering
    \includegraphics[width=1.0\linewidth]{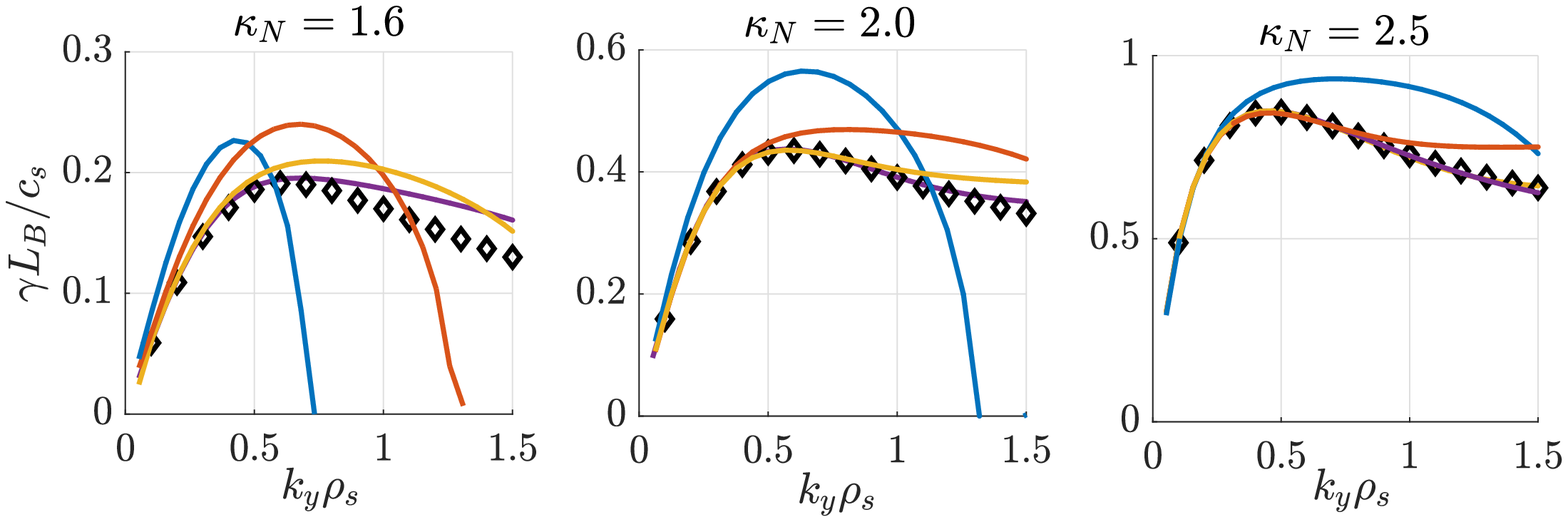}
    \vspace{-0.5cm}
    \caption{Growth rates of the linear entropy mode in the collisionless case ($\nu=0$) for three different drive values, $\kappa_N=1.6$ (left), $\kappa_N=2.0$ (middle) and $\kappa_N=2.5$ (right), keeping $\eta=0.25$. The growth rates are obtained with GENE with $N_{v_\parallel}=32$ and $N_\mu= 16$ velocity grid points (black diamonds) and different gyromoment sets: $(4,2)$ (blue), $(10,5)$ (red), $(20,10)$ (yellow) and $(30,15)$ (purple).}
    \label{fig:linear_analysis}
\end{figure}

 
%Obtained by neglecting the nonlinear terms in Eq. \eqref{eq:sapj_def}, the growth rate of the entropy mode is shown in \figref{fig:linear_analysis}.
Considering the results presented on \figref{fig:linear_analysis}, we first note that the gyromoment approach retrieves the converged results obtained with GENE, given a sufficiently large polynomial basis.
The convergence properties of the gyromoment model depend on the strength of the gradients and improve at steep gradients, confirming previous results obtained for the slab \citep{Frei2022LocalMode} and the toroidal ITG instability \citep{Frei2022Moment-BasedModel}.
Second, the results obtained with a number of polynomials below convergence show a stabilization of the high $k_y$ tail of the entropy mode and a larger peak growth rate.
Third, it is worth noting that, independently of the polynomial resolution, the growth rates obtained with a small number of polynomials agree with the converged results in the long wavelength limit, $k_y\ll 1$, highlighting the fact that the gyromoment method retrieves the fluid limit, even when a small set of gyromoments is used.
\par

%%%%%%%%%%%% NONLINEAR %%%%%%%%%%%%%%%%%%
 \subsection{Nonlinear collisionless simulations}
%% RESOLUTION PARAMETERS
Let us now consider the nonlinear case by including the $E\times B$ term in Eq. \eqref{eq:sapj_def}.
GENE results are used to benchmark our implementation (the GENE simulations presented here closely recall those by \cite{Hallenbert2022}).
We focus on three values of the density background gradient, $\kappa_N = 1.6$, $2.0$, and $2.5$, with $\eta = 0.25$ and $\nu=0$.
The system is evolved in a periodic box of dimensions $L_x\times L_y=120\times80$, for the lowest gradient value, and $L_x\times L_y=400\times 240$, for the highest gradient value.
In terms of spatial resolution, we consider a Fourier grid with $N=128$ and $M=32$ Fourier modes along the $x$ and $y$ directions, respectively, except for the steepest gradient case where we increase the resolution to $N=256$ and $M=128$ in order to reduce the need of artificial numerical dissipation.
The velocity space is represented by the Hermite-Laguerre basis $(P,J)=(4,2)$ extended up to $(P,J)=(20,10)$ at the lowest gradient.
GENE results are obtained using the same Fourier modes as the gyromoment simulation and a velocity grid resolution of $N_{v_\parallel}\times N_\mu=32\times  12$ points for the $(v_\parallel,\mu)$ velocity space in a box of dimension $L_{\vpar} \times L_\mu = 6\times 4$.
\modif{This ensures convergence of GENE results and that the results obtained by \cite{Hallenbert2022} are retrieved.
It is worth noting that \cite{Hallenbert2022} present the velocity space resolution we use as the minimum necessary not to compromise key results.
We confirm their claim by observing spurious predator-prey cycles when running lower resolution simulations at $\kappa_N=1.6$.}
\par
When running the collisionless cases, GENE uses a kinetic artificial diffusion term, $\nu_v (\Delta \vpar/2)^4\ddvpar^4 g_a$, with the diffusion parameter fixed to $\nu_v=0.2$ \citep{Pueschel2010OnMicroturbulence}.
Both codes use a spatial fourth-order hyperdiffusion term in both perpendicular directions $\mu_{HD} (k/k_{\mathrm{max}})^4$, with $0.5\leq \mu_{HD}\leq 5.0$, adjusted on the drive level in order to avoid energy pile-up without compromising the accuracy of results.
\modifii{For the intermediate values of the equilibrium gradient strength, we perform two simulations with GENE. 
The first simulation considers a constant level of numerical diffusion, while the second takes advantage of the adaptive numerical diffusion feature in GENE, as described in \cite{Hallenbert2022}}. 
This is done to ensure that the effect of the adaptive diffusion feature is not significant.
This test allows us to confirm that the level of transport is resilient to spatial hyperdiffusion. \\
While a comparison of the computational cost of the two approaches is not straightforward, we note that the number of gyromoments evolved is given by $N_{P,J}=(P/2+1)\times(J+1)$ (we take into account that only the even $p$ gyromoments are evolved in the Z-pinch geometry).
Therefore, 9 and 25 gyromoments are evolved in the $(P,J)=(4,2)$ and $(20,10)$ simulations, respectively.
This compares with the, approximately, $10^2$ velocity grid points used by GENE.
\par
%%%% TRANSPORT AVERAGE VALUE ANALYSIS
% \begin{figure}
%     \centering
%     \includegraphics[width=0.8\linewidth]{figures/transport_collisionless.eps}
%     \caption{Agreement plot of the saturated transport level, $\Gamma_x^{\infty}$, obtained with the gyromoment (GM) approach for $(P,J)=(4,2)$ (blue), $(10,5)$ (red) and $(20,10)$ (yellow) results and GENE $32\times 12$ in the collisionless case. The dashed black line shows the $x=y$ line.}
%     \label{fig:coll_transp_bench}
% \end{figure}
\begin{figure}
\centering
\begin{minipage}{.45\textwidth}
    \centering
    \includegraphics[width=1.0\linewidth]{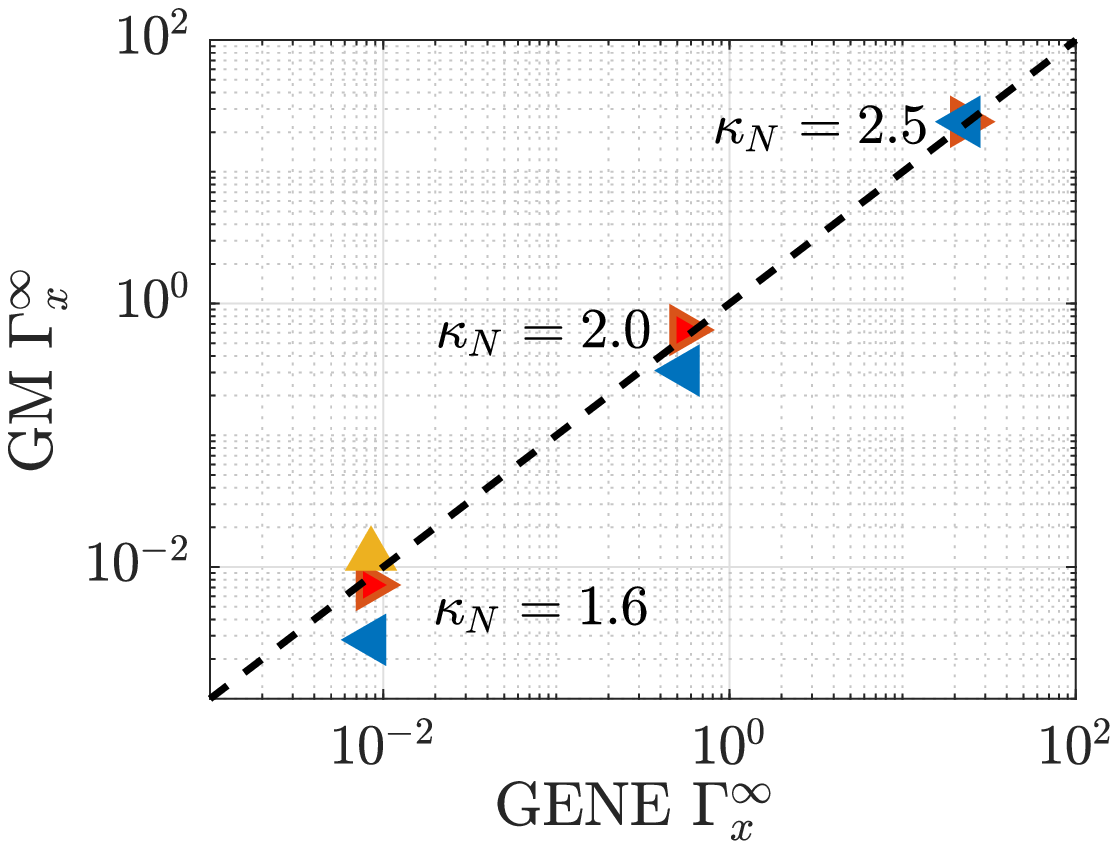}
     \vspace{-0.5cm}
   \caption{Comparison of the time-averaged transport level $\Gamma_x^{\infty}=\langle\Gamma_x\rangle_t$ obtained with the gyromoment (GM) approach for $(P,J)=(4,2)$ (blue), $(10,5)$ (red) and $(20,10)$ (yellow) and GENE, $\eta=0.25$. The time traces are presented on Fig. \ref{fig:turb_colless_transp}.}
    \label{fig:coll_transp_bench}
\end{minipage}
\qquad
\begin{minipage}{.45\textwidth}
    \centering
    \includegraphics[width=1.0\linewidth]{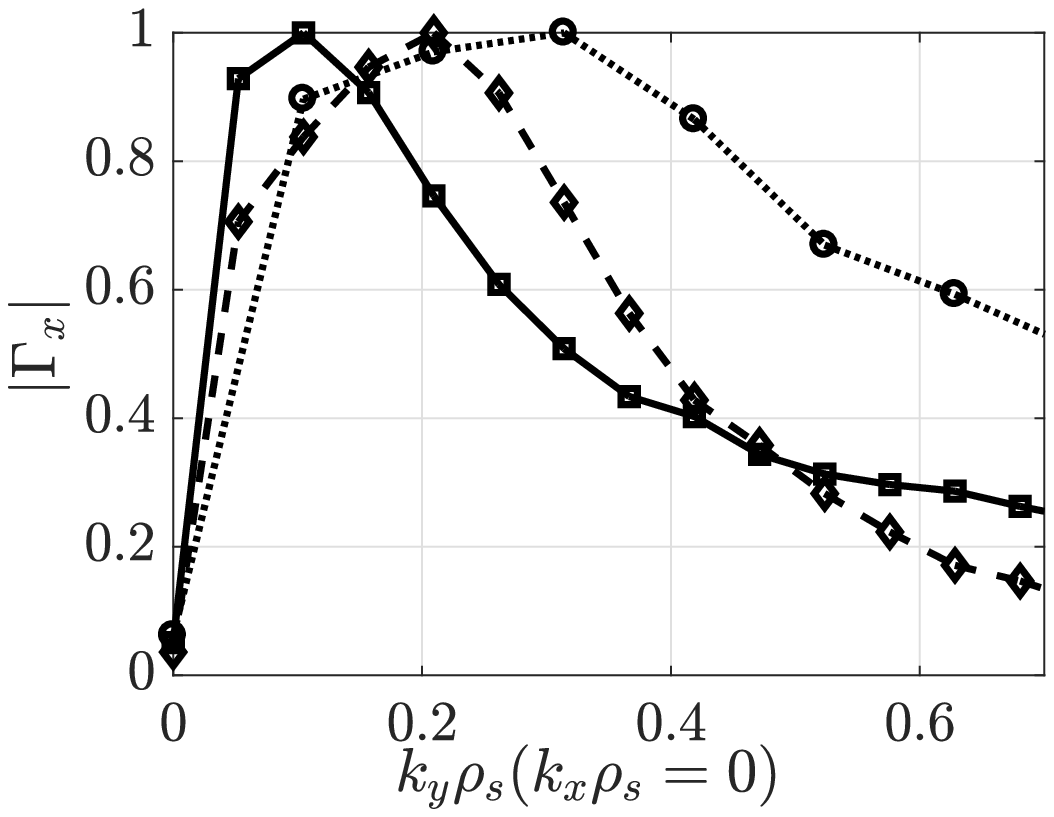}
    \vspace{-0.5cm}
    \caption{Spectrum of the radial particle flux, $\langle|\Gamma_x(k_x=0,k_y)|\rangle_t$ for the highest resolution simulations presented in \figref{fig:snapshots_colless} for $\kappa_N=2.5$ (solid squares), $\kappa_N=2.0$ (dashed diamonds) and $\kappa_N=1.6$ (dotted circles), $\eta=0.25$.}
    \label{fig:transport_spectrum}
\end{minipage}
\end{figure}
\par
\modif{We now focus on the quasi-steady turbulent state that is established after an initial transient following the initialization of the simulation.
In particular, we measure the saturated time-averaged turbulent transport level, $\Gamma_x^{\infty}$.
We observe that the gyromoment approach retrieves the saturated turbulent transport level obtained by GENE for all gradient values, given a sufficient number of gyromoments, over four orders of magnitudes (see \figref{fig:coll_transp_bench}).}
% Focusing on the turbulent state that is established after an initial transient following the initialization of the simulation, the gyromoment approach retrieves the saturated time-averaged turbulent transport level, $\Gamma_x^{\infty}=\langle \Gamma_x \rangle_t$, obtained by GENE for all drive values, given a sufficient number of gyromoments, over four orders of magnitudes (see \figref{fig:coll_transp_bench}).
As for the linear case, faster convergence with the number of gyromoments is observed in the case of the strongest gradient, with a set of $(P,J)=(4,2)$ gyromoments being sufficient for convergence.
This result might be surprising, considering the linear growth rate obtained with the same gyromoment resolution, significantly broader and showing a higher peak value than the converged value (see \figref{fig:linear_analysis}).
Even the results obtained with $(P,J)=(4,2)$ at the lowest gradient are surprisingly accurate when considering the accuracy of the linear growth rate.
\par
One can explain the faster convergence of the nonlinear simulations with respect to the evolution of the linear growth rate by considering the Fourier spectrum of the radial particle transport, $\langle\bm |\bm \Gamma \cdot \ex|\rangle_t$, at $k_x=0$ (see \figref{fig:transport_spectrum}).
It is found that transport is driven by fluctuations that occur on scale lengths that are larger than the ones at the peak growth rate of the entropy mode (see \figref{fig:linear_analysis}).
The peak of the transport spectrum shifts towards smaller wavelengths when the density gradient is reduced\modif{, in good agreement with the transport scaling, $\Gamma_x\sim \gamma ^2/k^3$, derived in Sec. 4}.
In addition, the turbulent quasi-steady state at low driving gradients is dominated by ZF which result from the growth of a KHI rising from $E\times B$ shear flow produced by the primary instability.
The growth rate of the KHI typically peaks at wavelengths that are twice as long as the primary instability \citep{Rogers2005}, thus pushing the dynamics towards larger spatial scales, where convergence of the gyromoment approach is achieved with a smaller number of gyromoments.
\par
Comparing the convergence of the entropy mode growth rate (\figref{fig:linear_analysis}), the convergence of saturated transport level (\figref{fig:coll_transp_bench}) and the spectrum of the radial particle transport (\figref{fig:transport_spectrum}), one can infer that the gyromoment simulations yield an accurate nonlinear transport level when the linear growth rate of the entropy mode is converged at the wavenumber of the transport spectrum peak.
For example, considering the $\kappa_N=2.0$ case, we note that the transport spectrum peaks at $k_y\simeq 0.3$.
The $(P,J)=(4,2)$ gyromoment result provides accurate growth rates for $k_y\lesssim 0.2$, thus yielding an inaccurate transport level.
On the other hand, the $(P,J)=(10,5)$ gyromoment set is linearly accurate for $k_y\lesssim 0.5$, which explains the correct saturated transport result.

% \begin{figure}
%     \centering
%     \includegraphics[width=0.6\linewidth]{figures/colless_transport_ky.eps}
%     \caption{Spectrum of the radial particle flux $|\Gamma_x(k_x=0,k_y)|$ averaged over time for the highest resolution simulations presented on \figref{fig:snapshots_colless} for $\kappa_N=2.5$ (blue), $\kappa_N=2.0$ (red) and $\kappa_N=1.6$ (yellow).}
%     \label{fig:transport_spectrum}
% \end{figure}
%%%%% TRANSPORT DYNAMICS ANALYSIS
\begin{figure}
    \centering
    \includegraphics[width=1.0\linewidth]{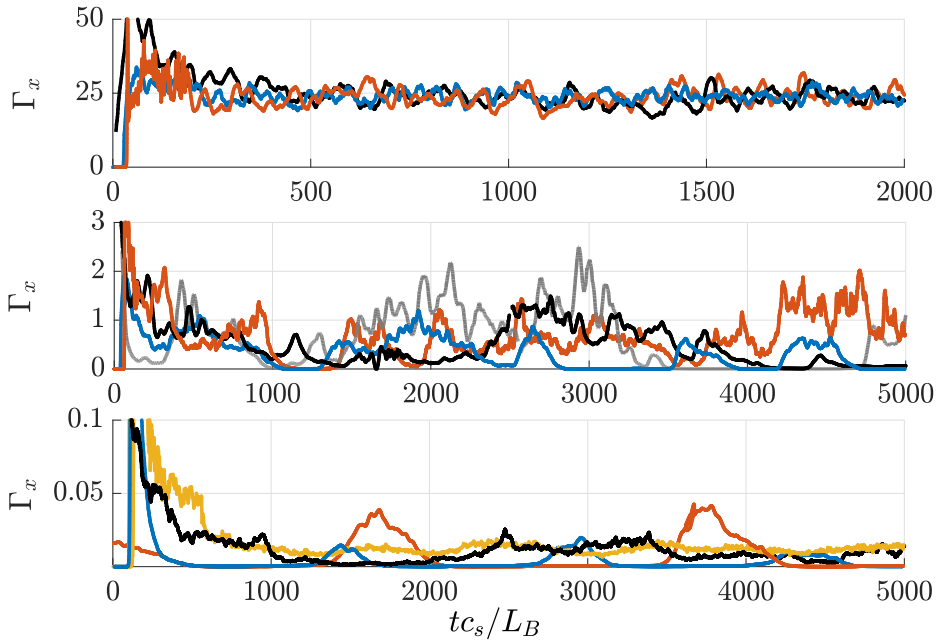}
    \vspace{-0.5cm}
    \caption{Radial particle transport $\Gamma_x(t)$ (see Eq. \eqref{eq:particle_flux}) from our nonlinear simulations.
    GENE results are obtained with constant $\mu_{HD}$ (black) and $\mu_{HD}$ set by an adaptive hyperdiffusion algorithm (gray). The gyromoment results are shown for $(P,J)=(4,2)$ (blue), $(P,J)=(10,5)$ (red) and $(P,J)=(20,10)$ (yellow). In all cases $\eta=0.25$.}
    \label{fig:turb_colless_transp}
\end{figure}

\begin{figure}
    \centering
    \includegraphics[width=1.0\linewidth]{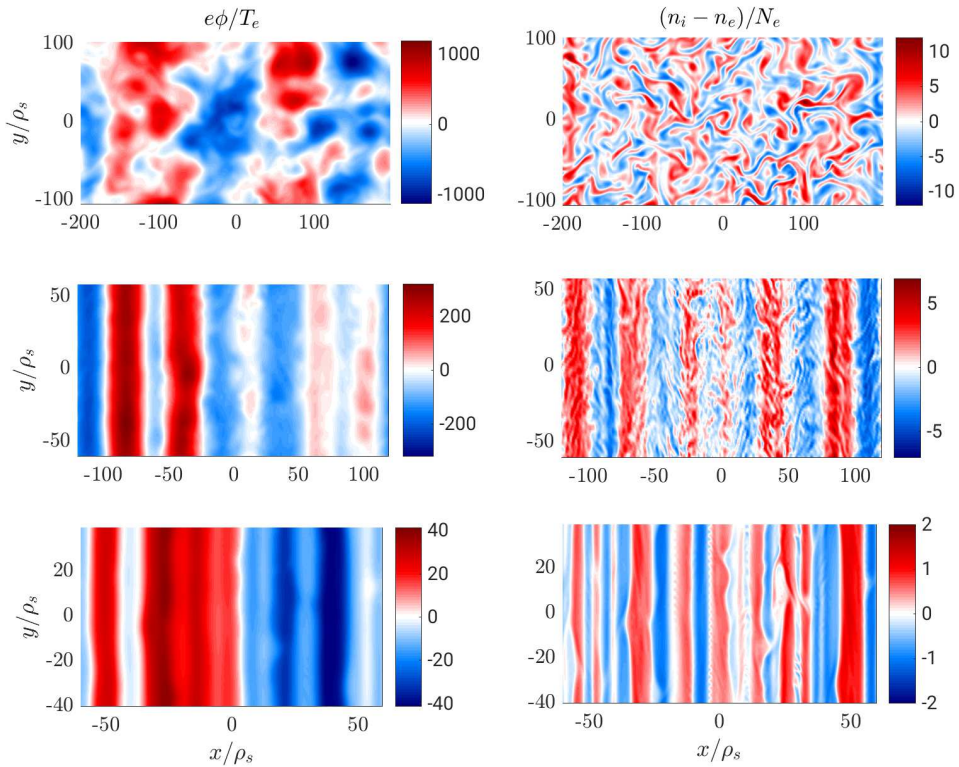}
    \vspace{-0.5cm}
    \caption{Snapshots of the electrostatic potential (left) and the charge density $n_i-n_e$ (right) in the collisionless case at the three drive values considered, i.e. $\kappa_N=2.5$ (top) with $(P,J) = (4,2)$, $\kappa_N=2.0$ with $(P,J) = (10,5)$ (middle) and $\kappa_N = 1.6$ with $(P,J) = (20,10)$ (bottom). In all cases, $\eta=0.25$.}
    \label{fig:snapshots_colless}
\end{figure}

For a finer analysis of our simulation results, we study the time dependence of the turbulent transport for the three equilibrium gradient values (see \figref{fig:turb_colless_transp}).
For instance, we note a negligible variation of the transport level with respect to the hyperdiffusion parameter, which mostly affects small-scale fluctuations, confirming \cite{Ricci2006a} and \cite{Hallenbert2022}.
% K_N = 2.5
At a large gradient level, $\kappa_N=2.5$, the gyromoment approach qualitatively and quantitatively agrees with GENE, showing an approximately constant transport.
The analysis of the turbulent eddies show fully developed turbulence, with a negligible role of ZF (see \figref{fig:snapshots_colless}).
% K_N = 2.0
At the intermediate gradient value, $\kappa_N=2.0$, time intervals characterized by a high turbulent transport level ($\Gamma_x\sim 1$) alternates with quiescent periods ($\Gamma_x\ll 1$), as shown in \figref{fig:snapshots_colless}.
The $(P,J)=(10,5)$ simulation is in good agreement with GENE results, while the $(P,J)=(4,2)$ results underestimate the average $\Gamma_x$ value because of longer low-transport intervals and lower burst level.
%!
However, our numerical tests show that the level of agreement between the gyromoment approach and GENE is within an uncertainty similar to the one related to the use of a constant hyperdiffusion or an adaptive numerical diffusion algorithm in GENE.
% K_N = 1.6
Finally, at the lowest gradient value considered, $\kappa_N=1.6$, the system is dominated by strong ZF that quench the turbulence reducing drastically the transport (see \figref{fig:snapshots_colless}).
As expected, the gyromoment method shows the largest discrepancies with respect to GENE in this case.
GENE simulation results in transport with small amplitude fluctuations occurring on long time scales around a plateau value, $\Gamma_x^{\infty}\sim 10^{-2}$.
On the other hand, the gyromoment approach shows bursts related to the damping of the ZF, for both the $(P,J)=(4,2)$ and $(10,5)$ resolutions.
Hence, even though the $(P,J)=(10,5)$ simulation results in an averaged transport level similar to GENE, an accurate description of the turbulent dynamics requires a larger number of gyromoments.
This is demonstrated by a $(P,J)=(20,10)$ simulation (see yellow line in \figref{fig:turb_colless_transp}), which agrees better with GENE results and does not produce the spurious bursts observed when a lower number of gyromoments is used.
\par
We note that bursts can also be obtained with GENE by reducing the $(\vpar,\mu)$ velocity grid resolution to $16\times 8$, keeping $\nu_v=0.2$.
Bursts are also obtained with a $32\times 16$ resolution when the velocity diffusion parameter is increased \modif{by a factor of $16$, i.e. $\nu_v = 3.2$, which ensures the same level of dissipation as in the coarser velocity resolution case.}
Thus, predator-prey cycles appear when a large level of diffusion is present in the velocity space.
This diffusion can also be introduced through simple collision models, such as the Lenard-Bernstein operator \citep{Lenard1958}.
Our results thus demonstrate that the effect of using a reduced number of gyromoments is comparable to the presence of diffusion in velocity space, with the level of diffusion that depends on the highest gyromoment considered.
% \begin{figure}
%     \centering
%     \includegraphics[width=1.0\linewidth]{figures/snapshot_turb_colless.eps}
%     \vspace{-0.5cm}
%     \caption{Electrostatic potential (left) and ion gyrocenter density (right) in the collisionless case at the three drive values considered, i.e. $\kappa_N=2.5$ (top) with $(P,J) = (4,2)$, $\kappa_N=2.0$ with $(P,J) = (10,5)$ (middle) and $\kappa_N = 1.6$ with $(P,J) = (20,10)$ (bottom). In all cases, $\eta=0.25$.}
%     \label{fig:snapshots_colless}
% \end{figure}
%Our nonlinear simulations thus demonstrate that the nonlinear turbulent state is less sensitive to convergence error than the linear growth rates of the primary instability.
%Indeed, we found good agreement in the level of transport between the gyromoments and GENE simulations even in cases where gyromoments' linear growth rates are not converged either in stability threshold nor in peak position.
%This is related to the fact that the transport spectrum and the nonlinear interactions are mainly due to large-scale fluctuations, as it is shown in \figref{fig:transport_spectrum}.
% \begin{figure}
%     \centering
%     \includegraphics[width=1.0\linewidth]{figures/transport_spectrum.png}
%     \caption{Amplitude of the radial transport modes $|\Gamma_x(k_x,k_y)|$ averaged over time of the highest resolution simulations presented on \figref{fig:coll_transp_bench} for $\kappa_N=2.5$ (top), $\kappa_N=2.0$ (middle) and $\kappa_N=1.6$ (bottom).}
%     \label{fig:transport_spectrum}
% \end{figure}
\par
\begin{figure}
    \centering
    \includegraphics[width=1.0\linewidth]{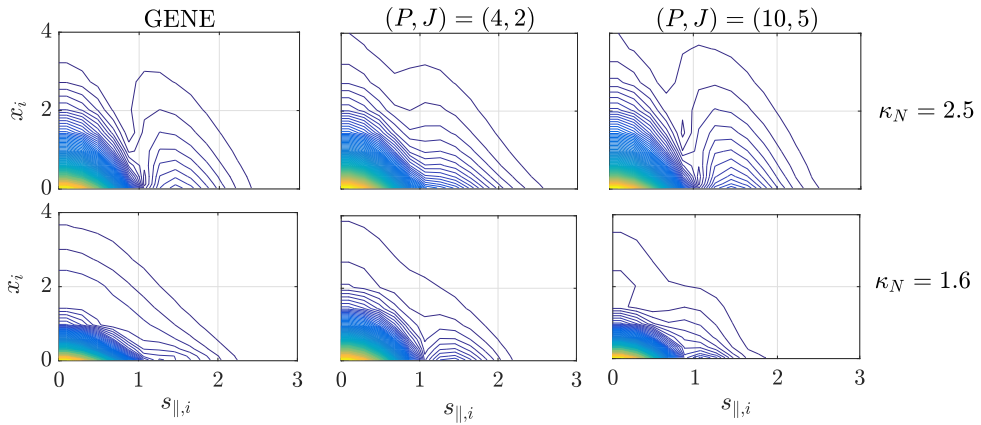}
    \vspace{-0.5cm}
    \caption{Time-averaged normalized ion velocity distribution function $|g_{v,i}(s_{\parallel,i},x_i)/g_{v,i}(0,0)|$. The results from GENE (left) and from the gyromoment approach $(P,J)=(4,2)$ (middle), $(P,J)=(10,5)$ (right), are presented for $\kappa_N=2.5$ (top) and $\kappa_N=1.6$ (bottom) keeping $\eta=0.25$ and $\nu=0$.}
    \label{fig:f_i}
\end{figure}

%%%%% KINETIC ANALYSIS
Since the representation of the velocity dependence of the distribution functions differs fundamentally between gyromoments and continuum approaches, we compare the time-averaged velocity distribution functions obtained by the gyromoments and GENE codes.
Within the gyromoments method, one can reconstruct the distribution function by using the gyromoments as coefficients of the Hermite-Laguerre basis.
This yields the averaged velocity distribution
$$
    g_{v,a}(\spara,x_a,t) = \sum_{p=0}^P \sum_{j=0}^J \langle N_a^{pj}(\bm k, t) \rangle_{k_x,k_y} H_p(\spara) L_j(x_a) \FaM.
$$
The results are presented in Figs. \ref{fig:f_i} for the ion distribution functions, considering $\kappa_N=1.6$ and $2.5$.
As for the transport properties, the agreement of the distribution functions between both codes depends on the gradient value.
At all gradient values considered, the $(P,J)=(4,2)$ gyromoment simulations lead to a smoothing of the distribution functions, reducing the sharp feature that appears around the thermal velocity (see \figref{fig:f_i}, around $\spara=1$).
This feature can be seen also in the lowest drive simulation in \figref{fig:f_i} where the $(P,J)=(10,5)$ gyromoment results present finer structures than the $(4,2)$ resolution.
This smoothing effect confirms the hypothesis that the use of a reduced number of gyromoments yields an effective diffusion in the velocity space.
%identified in the transport results, can be understood in more detail by considering how a generic kinetic diffusive term $\ddvpar^m g_a$ is expressed in the Hermite basis.
%Contrarily to the Fourier modes where $\ddx^m e^{ikx} \propto e^{ikx}$, the Hermite polynomials are not eigenfunctions of the diffusion operator, $\ddvpar^m H_p  \propto H_{p-m}$.
%Hence, in order to control kinetic diffusion of order $m$, one must solve the gyromoment hierarchy up to $P=m$.
%The same exercise can be done with the Laguerre basis and lead to a similar conclusion.

% \begin{figure}
%     \centering
%     \includegraphics[width=1.0\linewidth]{figures/f_a_contour_2c.eps}
%     \caption{Velocity distribution function for ions (top) and electrons (bottom) averaged over quasi-steady regime. The results of GENE (left), the GM approach for $(P,J)=(4,2)$ (middle) and $(P,J)=(10,5)$ (right) are presented for $\kappa_N=2.0$, $\eta=0.25$ and $\nu=0$.}
%     \label{fig:HP_fig2c_f}
% \end{figure}
\par
In conclusion, we remark that the gyromoment method shows its ability to simulate the Z-pinch nonlinear turbulent dynamics in the collisionless limit, which represents the most challenging regime for this approach.
Valid results are obtained at large gradient drives with a velocity space represented by only 9 gyromoments per species, compared to the, approximately, $ 10^2$ velocity grid points used in GENE simulations.
On the other hand, at the weakest gradient drive studied, convergence is obtained with a number of gyromoments approximately equal to the number of points used by GENE.
In all cases, results obtained with a lower number of gyromoments still provide a reasonable prediction of the time-averaged level of transport.
% Thus, the GM method shall be considered a versatile tool that can bridge the gap between reduced fluid models and high-fidelity ones.

% \begin{figure}
%     \centering
%     \includegraphics[width=1.0\linewidth]{figures/f_a_contour_2b.eps}
%     \caption{Time-averaged normalized velocity distribution function $|g_{v,a}(\spara,x_a)/g_{v,a}(0,0)|$ for ions (top) and electrons (bottom). The results from GENE (left) and the gyromoment approach $(P,J)=(4,2)$ (middle), $(P,J)=(10,5)$ (right), are presented for $\kappa_N=2.5$ , $\eta=0.25$ and $\nu=0$.}
%     \label{fig:HP_fig2b_f}
% % \end{figure}
% % \begin{figure}
%     \centering
%     \includegraphics[width=1.0\linewidth]{figures/f_a_contour_2a.eps}
%     \caption{Time-averaged normalized velocity distribution function $|g_{v,a}(\spara,x_a)/g_{v,a}(0,0)|$ for ions (top) and electrons (bottom). The results from GENE (left) and the gyromoment approach $(P,J)=(4,2)$ (middle), $(P,J)=(10,5)$ (right), are presented for $\kappa_N=1.6$ , $\eta=0.25$ and $\nu=0$.}
%     \label{fig:HP_fig2a_f}
% \end{figure}

% \newpage
\section{Collisional turbulent transport}\label{sec:collision}
\modif{Building on the benchmark of our gyromoment solver with the GENE code in the collisionless limit, we now study the gyromoment method at finite collisionality, in particular $\nu=0.1$ and $\nu = 0.01$. 
These values encompass the typical collision rate in the core of a tokamak device (e.g., the collision frequency estimate in the DIII-D cyclone base case corresponds to $\nu\sim 0.05$ in our normalized units \cite{Lin1999EffectsTransport}).
We first present the impact of collisions on the convergence of the Hermite-Laguerre basis using the Sugama collision operator.
Then, we investigate the properties of turbulence in a Z-pinch, as obtained by using different linear collision operators. }

\modif{
\subsection{Collisions and convergence}
Adding collisions to our system helps significantly the convergence of the moment approach.
Fig. \ref{fig:answer_lin_conv} shows the linear growth rates of the entropy mode for various Hermite-Laguerre basis, two collision frequency values, $\nu=0.01$ and $0.1$, and two gradient levels, $\kappa_N=1.6$ and $2.2$.
This illustrates that convergence is obtained at high collisionality and high gradient levels with a low number of polynomials.

\begin{figure}
    \centering
    \includegraphics[width=1.0\linewidth]{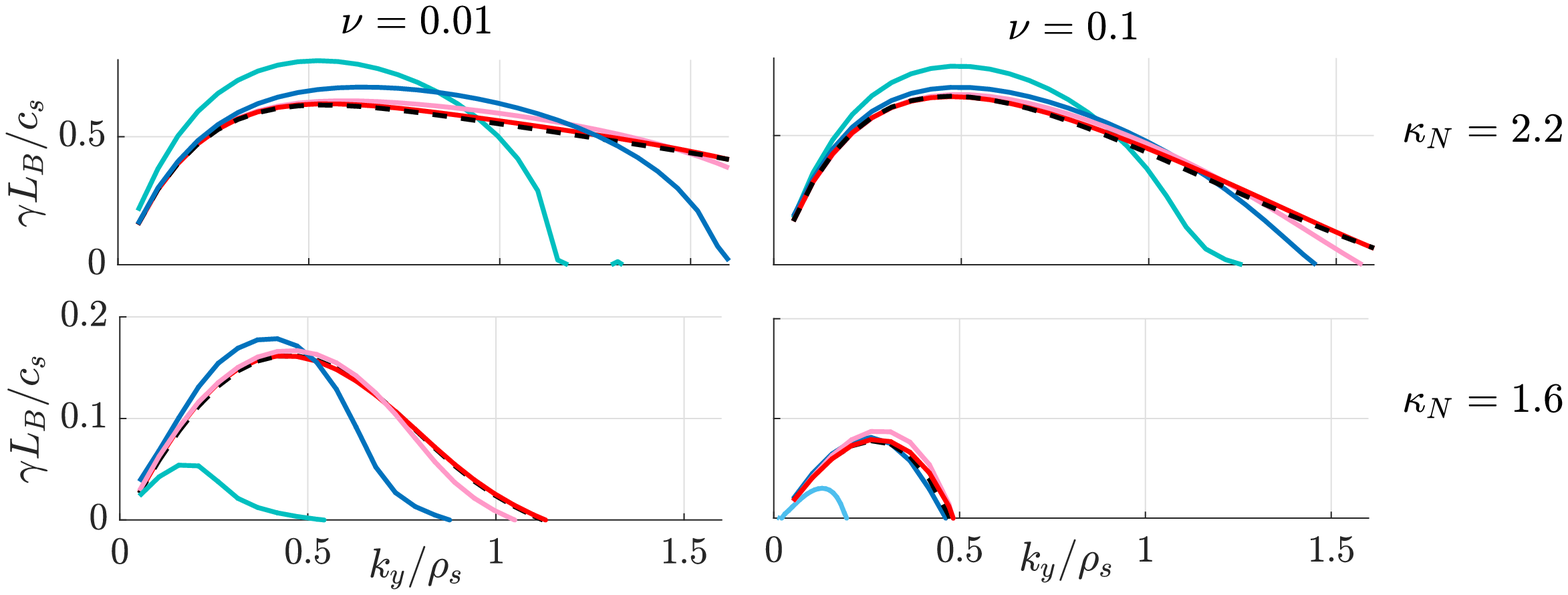}
    \vspace{-0.5cm}
    \caption{Convergence study of the entropy mode growth rate for $\kappa_N=2.2$ (top) and $\kappa_N=1.6$ (bottom) using the GK Sugama collision operator with $\nu = 0.01$ (left) and $\nu=0.1$ (right) for $\eta = 0.25$. The color indicates the polynomial basis used: $(P,J)=(2,1)$ (cyan), $(P,J)=(4,2)$ (blue), $(P,J)=(6,3)$ (pink), $(P,J)=(8,4)$ (red) and $(P,J)=(10,5)$ (black).}
    \label{fig:answer_lin_conv}
\end{figure}

Similarly, Fig. \ref{fig:answer_nonlin_conv} shows the nonlinear transport level for the parameters of Fig. \ref{fig:answer_lin_conv}.
One can observe, in particular, that the $(P,J)=(2,1)$ basis is sufficient at high collisionality and high gradient values.
In the other cases, nonlinear simulations carried out with this reduced polynomial basis overestimate the level of transport when the linear growth rate is overestimated if evaluated with the same number of polynomials, and vice versa.
Finally, the linear and nonlinear results presented in Figs. \ref{fig:answer_lin_conv} and \ref{fig:answer_nonlin_conv}, respectively, demonstrate that the basis $(P,J)=(4,2)$ is sufficient to obtain accurate results in the parameter region of interest.
Thus, the linear and nonlinear collisional simulations are performed using the polynomial basis $(P,J)=(4,2)$ in the following.
As an indication of the computational cost of our nonlinear simulations, we notice that one RK4 time-step for $(P,J)=(4,2)$ and $200\times 64$ spatial points is performed, on average, in $48$ms (wall clock time) when run on one Marconi node, i.e. $2\times24$-cores Intel Xeon 8160 (SkyLake) at 2.10 GHz.

\begin{figure}
    \centering
    \includegraphics[width=1.0\linewidth]{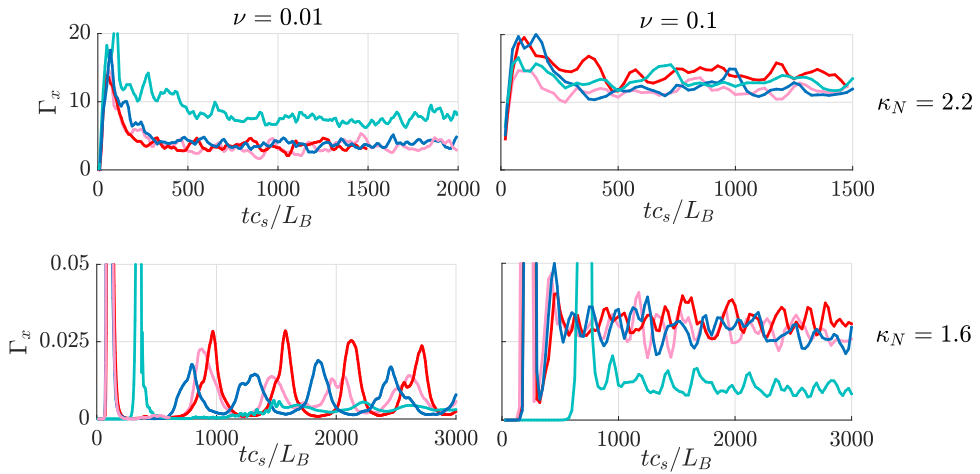}
    \vspace{-0.5cm}
    \caption{Convergence study of the turbulent transport time traces for $\kappa_N=2.2$ (top) and $\kappa_N=1.6$ (bottom) using the GK Sugama collision operator with $\nu = 0.01$ (left) and $\nu=0.1$ (right). The color indicates the polynomial basis used, $(P,J)=(2,1)$ (cyan), $(P,J)=(4,2)$ (blue): $(P,J)=(6,3)$ (pink), $(P,J)=(8,4)$ (red). The other parameters are $\eta = 0.25$ and $N_x=200$, $N_y=64$ for the spatial resolution.}
    \label{fig:answer_nonlin_conv}
\end{figure}

}

\subsection{Impact of collisions on the entropy mode and the Dimits shift}
% --------------- LINEAR RESULTS ----------------------
\par
\begin{figure}
    \centering
    \includegraphics[width=1.0\linewidth]{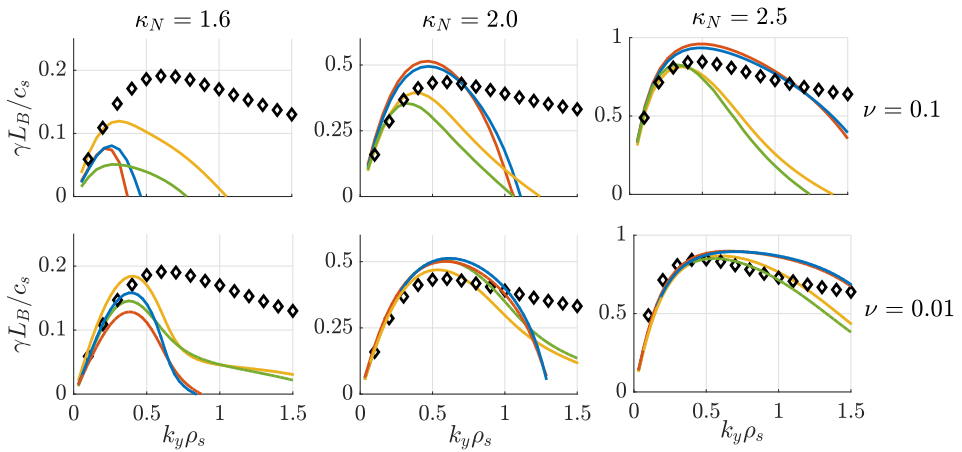}
    \caption{Linear growth rate of the entropy mode for different collision models and comparison with the collisionless results (black) for two different collision frequency, $\nu=0.1$ (top) and $\nu=0.01$ (bottom) and for three different drive values, $\kappa_N=1.6$ (left), $\kappa_N=2.0$ (middle) and $\kappa_N=2.5$ (right), keeping $\eta=0.25$. The different lines denote the Dougherty (red), Sugama (blue), Lorentz (yellow), and Coulomb (green) operators used in the gyromoment approach with a $(4,2)$ Hermite-Laguerre basis.}
    \label{fig:linear_coll}
\end{figure}

 Figure \ref{fig:linear_coll} shows the impact of collisions on the entropy mode linear growth rate for the cases considered in Sec. \ref{sec:colless}.
Collisions stabilize the tail of the entropy mode present at high $k_y$ in the collisionless regime because of diffusion in phase space, as observed in \cite{Ricci2006}.
This effect is recovered for both collision frequencies and by all the operators considered here, which also include gyrokinetic effects that induce strong damping for $k_y\gtrsim 1$.
%
% Collision operator comparison
% two families
At low $k_y$ and in the proximity of its peak value, one can observe that the growth rate is affected by collisions in different ways, depending on the collision model.
On the one hand, large-scale fluctuations are destabilized by collisional effects in the case of the Dougherty and Sugama collision operators for $\kappa_N=2.0$ and $\kappa_N=2.5$.
In this case, an increase of the growth rate at $k_y\sim 0.5$ is observed.
This effect is similar to the one observed in instabilities that have a fluid nature, such as the drift waves, which are destabilized by resistivity \citep{Goldston1995IntroductionPhysics}.
On the other hand, the collisional growth rate is smaller or close to the collisionless case when the Coulomb and Lorentz collision operators are used, for all values of $\kappa_N$ and $k_y$ considered.

% --------------- NONLINEAR RESULTS ----------------------
% \begin{figure}
%     \centering
%     \includegraphics[width=0.7\linewidth]{figures/collision_transport_w_g2k3.eps}
%     \caption{Collisional saturated transport level for different collision operators at $\nu=0.1$: Dougherty (red triangles), Sugama (blue squares), \textit{modified} Sugama (light blue squares), Coulomb (green diamonds) and Lorentz (yellow triangles). The collisionless results are also reported (black stars) with the mixing length estimate $\Gamma_x^\infty\sim \gamma_{p}^2$ (dashed black line). In all cases, $\eta=0.25$}
%     \label{fig:collision_transport}
% \end{figure}
\begin{figure}
    \centering
    \includegraphics[width=1.0\linewidth]{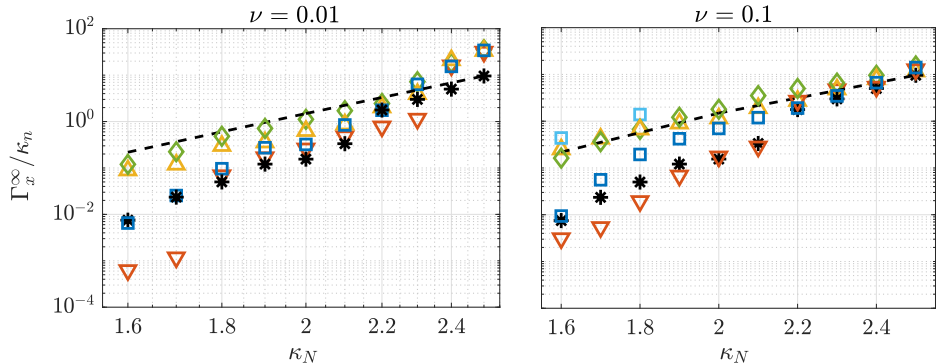}
    \vspace{-0.5cm}
    \caption{Collisional saturated transport level for different collision operators at $\nu=0.01$ (left) and $\nu=0.1$ (right): Dougherty (red triangles), Sugama (blue squares), \textit{modified} Sugama (light blue squares), Coulomb (green diamonds) and Lorentz (yellow triangles). The collisionless results are also reported (black stars) with the mixing length estimate $\Gamma_x^\infty\sim \gamma_{p}^2/k_p^3$ (dashed black line). In all cases, $\eta=0.25$}
    \label{fig:collision_transport}
\end{figure}

We now turn to the nonlinear results that include finite collisionality, and we discuss two scans of simulations, for $\nu=0.1$ and $\nu=0.01$, where the drive value is varied from $\kappa_N=1.6$ to $\kappa_N=2.5$, being $\eta = 0.25$.
The results are shown in \figref{fig:collision_transport}, where they are compared with the collisionless limit, which shows a Dimits threshold value $\kappa_N\simeq 2$, similarly to \cite{Hallenbert2022}, below which ZF suppress turbulence.
%
%Convergence tests show that collisions greatly reduce the number of gyromoments necessary for a correct estimate of the saturated transport level, as observed for the linear growth rate.
%This allows us to obtain accurate results using a reduced polynomial basis, $(P,J)=(4,2)$, for most of the results presented in \figref{fig:collision_transport}.
%For example, a $(P,J)=(10,5)$ gyromoment simulation provides approximately the same saturated transport level as a $(4,2)$ simulation for $\kappa_N=1.7$ with the Sugama operator.
%general comments
We observe that the effect of collisions vanishes at large drive values, where the ZF do not play a crucial role, in agreement with the observations in \cite{Ricci2006a}.
This suggests that the effect of collisions is mostly related to the ZF dynamics and, as we show later, through their damping and related weakening of the associated transport barrier.
\par
When turbulence is fully developed, the amplitude of the fluctuations can be estimated considering a balance between the nonlinear saturating terms and the linear drive, $\ddt \sim \bm v_{E\times B}\cdot \grad$.
\modif{
This yields $\gamma \sim k^2 \phi$ and, thus, $\phi\sim \gamma/k^2$, considering the peak linear growth rate, $\gamma$, the associated wavenumber $k$, and with the assumption of circular eddies ($k_x \sim k_y \sim k$).
Using Poisson equation, one observes that the particle density scales with the potential fluctuations, $n\sim \phi$, which leads to the estimate of the radial particle transport $\Gamma_x\sim \gamma ^2/k^3$.
}
This scaling, based on the collisionless peak value of the entropy mode instability, is shown in \figref{fig:collision_transport}, revealing that it captures well the dependence of $\Gamma_x$ at strong gradients, where the effect of ZF is weak.
On the other hand, the reduction of the transport by the ZF cannot be captured by this mixing-length estimate at a low gradient value.

\par
At medium and low levels of the driving gradient, where ZF are expected to play a role, according to the collisionless values, the different collision models lead to significantly different results\modif{, for both the $\nu = 0.01$ and $\nu=0.1$ cases}.
In particular, the Sugama and Dougherty tend to differ from the Lorentz and Coulomb operators.
The difference cannot be explained solely in terms of linear growth rate since the Coulomb operator linear results differ from the Lorentz results at lower gradient values (see \figref{fig:linear_coll}).
In fact, the ZF quenching of the turbulence \citep{Kobayashi2012} has a strong dependence on the collision model.
\par
\modif{Decreasing the collision frequency by a factor of ten, i.e. between $\nu=0.1$ and $\nu=0.01$ (see \figref{fig:collision_transport}), reduces the gap between collisional and collisionless results in the majority of parameters and collision models studied.
However, it is worth noting that, at a high gradient value, the transport does not approach the collisionless value monotonically with resistivity.
This phenomenon is due to a combination of the tertiary instability affecting the ZF and the damping of turbulence due to collisions. 
In fact, \cite{Ricci2006a} observed a non-monotonic dependence of transport to collisionality at large gradient values as well.
We note that this feature does not depend on the chosen collision model.}
\par
%% Dougherty analysis
The results obtained with the Dougherty operator appear to most closely approach the collisionless case, with Dougherty being the only operator that shows a Dimits shift at $\kappa_N\simeq 2.1$ for $\nu = 0.1$ and $\kappa_N \simeq 2.3$ for $\nu = 0.01$.
This similarity can be explained by the simplicity of the Dougherty model, which is mainly composed of kinetic and spatial diffusion terms that are present, albeit at smaller amplitude and for numerical reasons, also in the collisionless case.
Concerning the collisionless case, we expect that the slight reduction of transport at the lowest drive level is due to the reduced linear drive.
\modif{When reducing the collisionality at $\kappa_N\leq1.7$, we observe that transport is significantly reduced in comparison to the collisionless case.
In this regime, the ZF are stable and yield small bursts of transport occurring over large time intervals, approximately $2000 L_B/c_s$.
We report that increasing the polynomial basis to $(P,J)=(8,4)$ does not affect significantly the result obtained with the $(P,J)=(4,2)$ basis at these lower gradient values.}
\par
\begin{figure}
    \centering
    \includegraphics[width=1.0\linewidth]{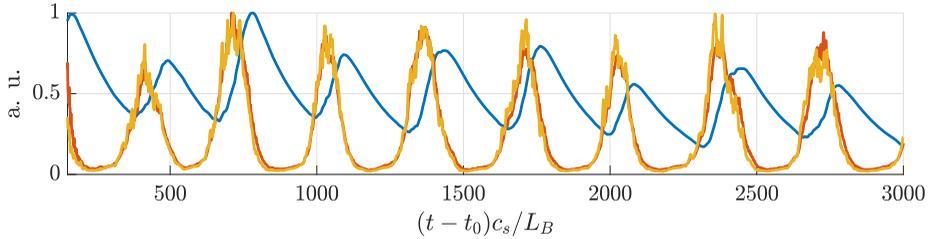}
    \vspace{-0.5cm}
    \caption{Amplitude of the normalized spectral energy for zonal modes ($\sum_{k_y=0}|\phi_k|$; blue), non-zonal modes ($\sum_{k_y\neq 0}|\phi_k|$; red) and transport level ($\Gamma_x$; yellow) obtained for a simulation with the Sugama collision operator for $\kappa_N=1.6$, $\eta=0.25$ and $\nu = 0.01$, with $(P,J)=(4,2)$ gyromoments.}
    \label{fig:pred_prey}
\end{figure}

%% Sugama analysis
Similarly to the Dougherty operator, the Sugama operator yields a regime of suppressed transport at low gradient values and a regime of fully developed turbulence at large gradient values.
However, at intermediate gradient levels, transport is remarkably larger in comparison to the collisionless results and Dougherty operator, with the oscillations between quiescent and turbulent periods (see \figref{fig:coll_transp_bench}) being replaced by fluctuations around a plateau value with persistent ZF structures.
This feature, also observed with the Lorentz and Coulomb operators, can be explained by a ZF damping sufficiently strong to continuously allow fluctuations to grow in the ZF regions where the $\bm E\times \bm B$ velocity shear vanishes \citep{Ivanov2020ZonallyTurbulence}.
In the context of the predator-prey cycles, this case corresponds to an overlap of bursts.
This effect is reduced when the collision frequency is decreased to $\nu=0.01$ where cyclic transport dynamics, previously identified by \cite{Kobayashi2015a} and shown in \figref{fig:pred_prey}, are obtained.
The frequency of these bursts is directly related to the ZF damping rate due to the collision operator that dissipates the ZF structures (highlighted by the decreasing phase of the zonal energy, blue line of \figref{fig:pred_prey}), and the primary instability growth rate (underlined by the slope of the increasing part of the non-zonal energy, red line in \figref{fig:pred_prey}).
\modif{The fact that the burst period is shorter with the Sugama operator than the Dougherty operator indicates that a strong ZF damping mechanism resides in the higher gyromoments coupling present in the Sugama operator.}
% It is worth noting that the period of bursts obtained is of the order of $T\sim 500 L_B/c_s$.
% For typical fusion plasma parameters, $T_e \sim 2$ KeV and $L_B=1$ m \citep{Greenfield1997EnhancedDIII-D}, the period in physical units becomes $T\sim 1$ [ms] which is close to the bursting behavior reported in TFTR core plasmas \citep{Mazzucato1996TurbulentShear,Lin1999EffectsTransport}.
\par
% Lorentz
The reduction of the transport level with respect to the mixing length estimate at the lowest gradients is less pronounced with the Lorentz operator than with the Dougherty and Sugama operators.
The difference between the Sugama and the Lorentz model is mainly due to the energy diffusion term contained in the field part of the former collision operator.
In fact, the pitch-angle scattering Lorentz operator does not contain any energy diffusion term, while the Sugama model uses an ad-hoc energy diffusion term in the field part of the collision operator (on the other hand, the spatial diffusion terms of the Lorentz and Coulomb operators coincide).
Confirming the importance of having an accurate description of the energy diffusion, we note that tests at low drive values, where we modify the Sugama operator by zeroing out the ad-hoc energy diffusion term \modif{(while also breaking the Sugama conservation properties)}, show a significant increase of the transport level (light blue squares in \figref{fig:collision_transport}).
It is worth noting that no bursts are observed in Lorentz simulations even in the low collisionality case.
\par
% Coulomb
The Coulomb collision operator simulations do not show remarkable differences in comparison to the Lorentz collision operator in the high collisionality case.
Both operators maintain a high level of transport, even at low gradient values.
It is worth noting that the Coulomb collision operator induces the largest level of transport than all other collision operators for almost every $\kappa_N$, which can be surprising since the related linear growth rate is smaller than the one yielded by the other collision operators (see \figref{fig:linear_coll}).
\modif{In particular, in the low collisionality case the Coulomb operator maintains a high transport level also with respect to the Lorentz operator.}
\par
\par
\begin{figure}
    \centering
    \vspace{-0.5cm}
    \includegraphics[width=1.0\linewidth]{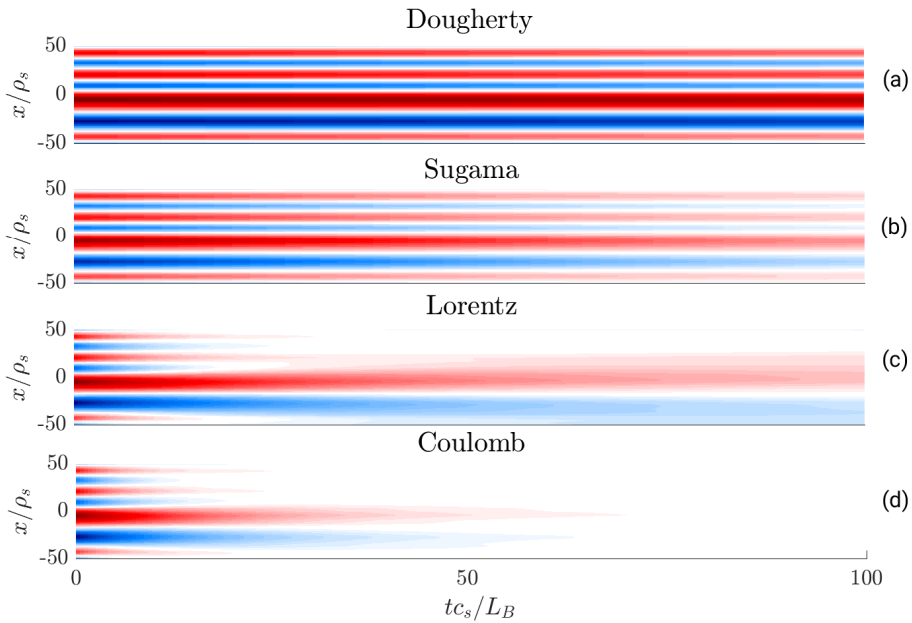}
    \caption{Time evolution of the $y$-averaged zonal flows profile, $\langle\ddx\phi\rangle_y$, for the Dougherty (a), Sugama (b), Lorentz (c) and Coulomb (d) collision operators, using the saturated state of the collisionless simulation at $t_0=5000$ for $\kappa_N=1.6$, $\nu=0.1$ and $\eta=0.25$ as initial conditions.}
    \label{fig:zf_damping_st}
\end{figure}

% --------------- ZF damping tests ----------------------
Confirming that collisions regulate transport through the ZF damping, we now describe a detailed study of this mechanism, as induced by the different collision models.
We consider the nonlinear collisionless saturated states for $\kappa_N = 1.6$, 2.0, and 2.5 (see \figref{fig:turb_colless_transp}) as initial conditions for a set of simulations that use different collision models.
We isolate the damping effect by removing the entropy mode drive, $\kappa_N=0$, and we use a $(P,J)=(4,2)$ gyromoment set with $\nu=0.1$.
We let the system evolve and follow the damping of the ZF profile.
The results of this numerical experiment can be first observed qualitatively in \figref{fig:zf_damping_st} where the averaged radial ZF profile, $\langle\ddx\phi\rangle_y$, is plotted as a function of time for each collision operator considered.
\figref{fig:zf_damping_st} reveals that the effect of collisions on the ZF profile is highly dependent on the operator model.
The Dougherty model does not significantly affect the ZF structure, while the Sugama operator leads to their damping.
The Lorentz operator filters the initial ZF structure, decreasing the amplitude of short wavelength ZF, while a long wavelength mode survives.
Finally, the Coulomb operator strongly damps the ZF at all wavelengths.
Thus, confirming our hypothesis that the different ZF damping is responsible for the different level of transport, the smallest transport values observed on \figref{fig:collision_transport} correspond to the operators that allow the smallest scale of the ZF structure to survive.
\par
\begin{figure}
    \centering
    \includegraphics[width=1.0\linewidth]{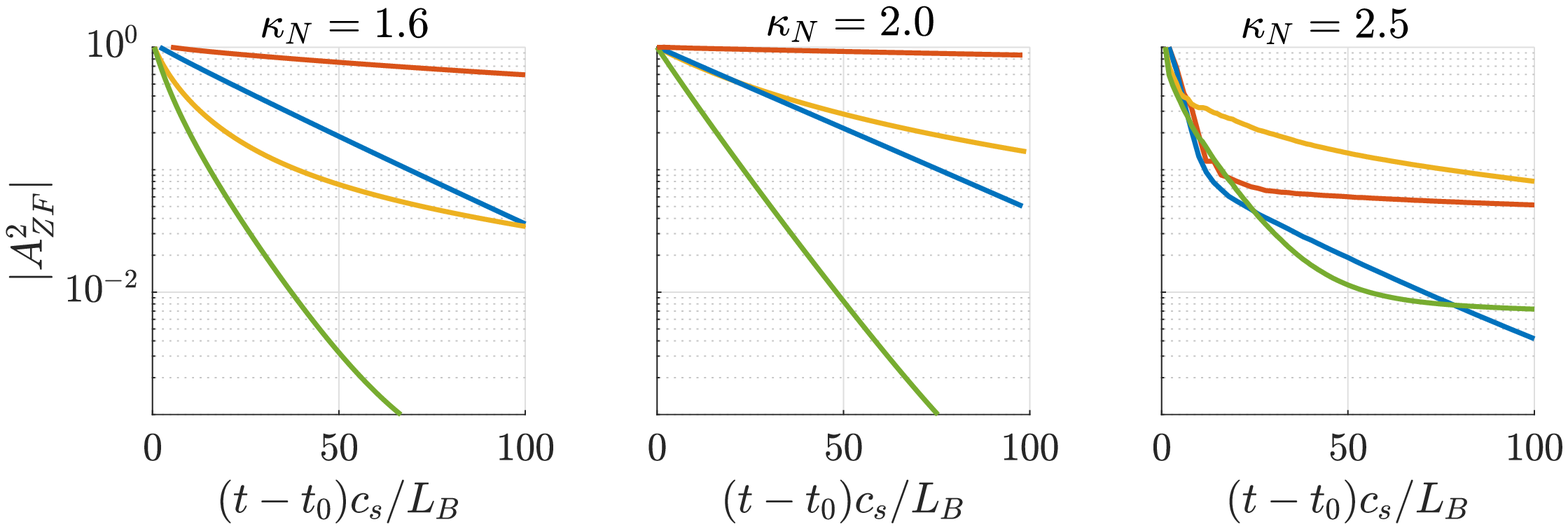}
    \vspace{-0.5cm}
    \caption{Time evolution of $A_{ZF}^2$ (see eq. \ref{eq:zonal_wave_amp}) for the Dougherty (red), Sugama (blue), Lorentz (yellow), and Coulomb (green) collision operators used in the gyromoments approach with a $(4,2)$ Hermite-Laguerre basis. The ZF initial conditions are the ones obtained from the $\kappa_N=1.6$ (left), $\kappa_N=2.0$ (middle), and $\kappa_N=2.5$ (right) collisionless simulations. In all cases $\eta=0.25$.}
    \label{fig:zf_damping_energy}
\end{figure}

As a further confirmation and a more quantitative analysis of the results shown in \figref{fig:zf_damping_st}, we define the normalized \modif{ZF energy}, i.e.
\begin{equation}
    A^2_{ZF}(t) =\frac{\int \mathrm dx \langle\phi\rangle_{y}^2(t)}{\int \mathrm dx \langle\phi\rangle_{y}^2(0)},
    \label{eq:zonal_wave_amp}
\end{equation}
and study its time evolution for each collision operator in \figref{fig:zf_damping_energy}.
As initial conditions, we consider the ZF obtained in the $\kappa_N=1.6$, $\kappa_N=2.0$, and $\kappa_N=2.5$ collisionless simulations.
Focusing on the damping at early times, $\partial_t A^2_{ZF}|_{t=0}$ (the growth of the linear instability alters the ZF damping at timescales $1/\gamma \sim 10$), this analysis unveils a clear difference between Dougherty and Sugama operators, while these operators provide very similar linear growth rates.
We also observe that the Lorentz and Coulomb operators yield similar damping, corresponding to a similar transport level in the nonlinear simulations.
% Thus, we show that the ZF damping is a better candidate to estimate the saturated transport level than the linear growth rate in the cases we considered.
Thus, we can deduce that, unlike the linear growth rate, the saturated transport level is directly related to the ZF damping rate.

% \newpage
\section{Conclusions}\label{sec:conclusion}
In the present paper, the first nonlinear gyrokinetic simulations carried out using a gyromoment approach and including advanced collision models are presented.
By implementing the moment hierarchy in Eq. \eqref{eq:gmhierarchy}, turbulence in a two-dimensional Z-pinch geometry is studied.  
\par
We first present a benchmark with the continuum GK code GENE that demonstrates the ability of the gyromoment approach to simulate accurately the nonlinear evolution of the entropy mode, even in the collisionless limit.
We show that the convergence behavior of the nonlinear results follows the same trend as the linear ones, i.e. convergence properties improve with the increase of the gradient strength.
However, accurate nonlinear results require only that the linear growth rate of the modes developing at large scales are accurately resolved.

We then extend the nonlinear results, adding collisions with the use of four different collision operator models.
We observe that the gyromoment simulations converge with a lower number of gyromoments than in the collisionless case.
With a Dimits threshold identified around $\kappa_N\sim2$ in the collisionless case, the influence of collisions on the transport level \modif{becomes particularly evident} for $\kappa_N<2$.
\modif{This confirms previous studies \citep{Lin1999EffectsTransport,Ricci2006a,Ricci2010}, pointing} that collisional effects are mainly related to the dynamics of the ZF.
Our results highlight the disagreement between Dougherty, Sugama, Lorentz, and Coulomb GK collision models, in the linear growth rate and, even more, in the level of nonlinear transport.
We show that the analysis of the linear results is not sufficient to predict the difference observed in the saturated transport level.
However, we observe a direct link between ZF damping and transport level, which could be used to develop a reduced model of the transport level in a future work.
\modif{By demonstrating for the first time that the transport level in ZF dominated regime is highly dependent on the collision model in use, we point out that the choice of the collision operator should be properly considered in GK turbulence simulations where ZF are present.}

\modifii{In a more general context, the present study is a \modif{first} step towards the nonlinear simulation of the tokamak boundary based on the use of the gyromoment approach.
Our plan is to consider nonlinear simulations in the $s-\alpha$ flux-tube geometry as a next step, expanding the linear study presented in \cite{Frei2022Moment-BasedModel}.
Using the cyclone base case as a reference \citep{Dimits2000ComparisonsSimulations}, we will have a benchmark for evaluating the performance of the gyromoment approach.
In particular, the ability of the Hermite-Laguerre approach to accurately resolve nonlinear trapped-particle dynamics remains an open area of research.
%as well as the impact of various collision models on such systems.
We will then turn to the simulation of the tokamak boundary.
%
% It is worth noting that \cite{Frei2022Moment-BasedModel} also uses the gyromoment approach to investigate edge-related instabilities, such as kinetic ballooning modes, laying the foundation for future nonlinear simulations in the tokamak edge conditions.
According to \cite{Frei2022Moment-BasedModel}, we anticipate that the high pressure gradient and level of collisionality present in the tokamak edge will improve the convergence of the nonlinear gyromoment hierarchy with respect to core conditions.
However, the $\delta f$ assumption will need to be relaxed when simulating the SOL.
%, however it is important to note that increasing the gradient strength indefinitely may eventually break the $\delta f$ assumptions.
The convergence behavior of a Hermite-Laguerre moment approach in a full-F nonlinear framework and the influence of its closure model remains a central question.
}

\section*{Acknowledgements}
The authors acknowledge helpful discussions with A. Cerfon, S. Brunner, J. Ball, L. Villard, A. Hallenbert, A. Vol\v{c}okas, L. Driever, and L. Simons.
This research has been carried out within the framework of the EUROfusion Consortium and has received funding from the European Union via the Euratom Research and Training Programme (Grant Agreement No101052200 — EUROfusion).
The views and opinions expressed herein do not necessarily reflect those of the European Commission.
The simulations presented herein were carried out in part on the CINECA Marconi supercomputer under the TSVVT421 project and in part at CSCS (Swiss National Supercomputing Center).
This work was supported in part by the Swiss National Science Foundation.

\modifii{
\section*{Declaration of interests}
The authors report no conflict of interest.
}

% \subsubsection*{Acknowledgement}

%%%%%%%%%%%% APPENDIX %%%%%%%%%%%%%%%%%%
\newpage
\appendix
\modif{
\section{Truncated gyromoment hierarchy and comparison with gyrofluids and reduced fluid models}
\label{appendix:EHW}
This appendix focuses on the link between the gyromoment model and other moment-based models, in particular, the gyrofluid models of \cite{Brizard1992NonlinearPlasmas} and the further simplified models of \cite{Hasegawa1978Pseudo-three-dimensionalPlasma}, \cite{Hasegawa1983} and \cite{Dewhurst2009TheTurbulence}.
We write the two-dimensional Z-pinch truncated gyromoment hierarchy, Eq. \eqref{eq:gmhierarchy}, explicitly for the Hermite-Laguerre basis $(P,J)=(4,2)$ in the collisionless limit and zero truncation closure.
We then identify the set of gyromoments required to obtain the gyrofluid model of \cite{Brizard1992NonlinearPlasmas} in the same geometry.
Finally, we take the long wavelength, cold-ion, drift-kinetic limit to obtain a single vorticity equation similar to the extended Hasegawa-Mima model.

\subsection{Truncated reduced moment hierarchy in a two-dimensional Z-pinch}
The collisionless gyromoment hierarchy in a Z-pinch, Eq. \eqref{eq:gmhierarchy}, writes explicitly
\begin{align}
    &\ddt N_a^{pj}+\sum_{n=0}^{\infty}\curlyparenthesis{\sum_{s=0}^{n+j}d_{njs} N_a^{ps},\kernel_n\phi}+ \frac{\tau_a}{z_a} i k_y \squareparenthesis{(2j+1)n_a^{pj} - (j+1)n_a^{p,j+1}-jn_a^{p,j-1}}\nonumber\\
    % %
    &+ \frac{\tau_a}{z_a} i k_y \squareparenthesis{\sqrt{(p+1)(p+2)} n_a^{p+2,j} + (2p+1)n_a^{p,j} + \sqrt{p(p-1)}n_a^{p-2,j}}\nonumber\\
    &=\kappa_N ik_y\phi[\kernel_j\delta_{p0}+\eta\kernel_j\frac{\sqrt{2}}{2}\delta_{p2} + \eta(2j\kernel_j-[j+1]\kernel_{j+1}-j\kernel_{j-1})\delta_{p0}].
    \label{eq:GK_moment_hierarchy}
\end{align}
% The most interested reader may want to unfold the hierarchy for the $(P,J)=(4,2)$ basis.
Defining $D_t (N_a^{pj})= \ddt N_a^{pj} + \sum_{n=0}^{\infty}\curlyparenthesis{\sum_{s=0}^{n+j}d_{njs} N_a^{ps},\kernel_n\phi}$ and considering only even-$p$ gyromoments due to the Z-pinch symmetry, the truncated hierarchy up to $(P,J)=(4,2)$ writes
\newcommand{\Dt}[1]{D_t (#1)}
\newcommand{\truncate}[1]{}
\begin{align}
    &\Dt{N_a^{00}} + \frac{\tau_a}{z_a} i k_y \squareparenthesis{2 N_a^{00} + \sqrt{2} N_a^{20} - N_a^{01}}  
    % \nonumber\\&\qquad \qquad
    =  \squareparenthesis{\roundparenthesis{\kappa_N - 2}\kernel_0 +  ( 1- \kappa_T) \kernel_1} ik_y \phi     \label{eq:momenthierarchy_start}\\
    &\Dt{N_a^{20}} + \frac{\tau_a}{z_a} i k_y \squareparenthesis{ \sqrt{2} N_a^{00} + 6 N_a^{20} + \sqrt{12}N_a^{40} - N_a^{21}}
    % \nonumber \\ &
    = \frac{\sqrt{2}}{2} (\kappa_T-2)  \kernel_0 ik_y  \phi\\
    &\Dt{N_a^{01}} + \frac{\tau_a}{z_a} i k_y \squareparenthesis{- N_a^{00} + 4 N_a^{01} + \sqrt{2} N_a^{21} - 2N_a^{02}  }
    \nonumber \\ &\qquad \qquad
    =  \squareparenthesis{\roundparenthesis{1-\kappa_T}\kernel_0 +  (\kappa_N+2\kappa_T-4)\kernel_1 + (2-2\kappa_T)\kernel_2} ik_y \phi\\
    &\Dt{N_a^{40}} + \frac{\tau_a}{z_a} i k_y \squareparenthesis{ \sqrt{12} N_a^{20} + 10 N_a^{40} \truncate{+\sqrt{30} N_a^{60}} - N_a^{41}} = 0 \\
    &\Dt{N_a^{21}} + \frac{\tau_a}{z_a} i k_y \squareparenthesis{- N_a^{20} + \sqrt{2} N_a^{01} + 8 N_a^{21} + \sqrt{12} N_a^{41}- 2N_a^{22} }
    % \nonumber \\ &\qquad \qquad
    = \frac{\sqrt{2}}{2}\roundparenthesis{\kappa_T-2}\kernel_1ik_y\phi \\
    &\Dt{N_a^{02}} + \frac{\tau_a}{z_a} i k_y \squareparenthesis{-2 N_a^{01} + 6 N_a^{02} + \sqrt{2} N_a^{22} \truncate{-3 N_a^{03}}  } = \squareparenthesis{2\kernel_1-6\kernel_2\truncate{-3\kernel_3}}ik_y\phi      \label{eq:momenthierarchy_21_end}\\
    &\Dt{N_a^{41}} + \frac{\tau_a}{z_a} i k_y \squareparenthesis{- N_a^{40} + \sqrt{12} N_a^{21} + 12 N_a^{41} \truncate{+ \sqrt{30} N_a^{61}} - 2N_a^{42} } = 0 \\
    &\Dt{N_a^{22}} + \frac{\tau_a}{z_a} i k_y \squareparenthesis{- 2N_a^{21}+ \sqrt{2} N_a^{02} + 10 N_a^{22} + \sqrt{12} N_a^{42} - 3N_a^{23} } = -\sqrt{2}\kernel_2ik_y\phi \\
    &\Dt{N_a^{42}} + \frac{\tau_a}{z_a} i k_y \squareparenthesis{- 2N_a^{41} + \sqrt{12} N_a^{22} + 14 N_a^{42} \truncate{+ \sqrt{30} N_a^{62}} \truncate{- 3N_a^{43}} } = 0 \label{eq:momenthierarchy_42_end}
 \end{align}
 In Eqs. (\ref{eq:momenthierarchy_start}-\ref{eq:momenthierarchy_42_end}), the moments $N_a^{60},N_a^{03},N_a^{61},N_a^{23},N_a^{62}$ and $N_a^{43}$ are set to vanish by truncating the moment hierarchy.
Consequently, the Poisson equation is now a truncated version of Eq. \eqref{eq:poisson_moments}, i.e.
\begin{equation}
    \squareparenthesis{\sum_a \frac{z_a^2}{\tau_a}\roundparenthesis{1-\sum_{n=0}^{J}\kernel^2_n}}\phi = \sum_a z_a \sum_{n=0}^{J}\kernel_n N_a^{0n}.
    % \label{eq:poisson_moments}
\end{equation}

\subsection{Comparison with gyrofluid model}
Direct comparison with the gyrofluid model in \cite{Brizard1992NonlinearPlasmas} is obtained by expressing the gyrofluid moments ($n_a,\upara,\Pperpa,\Ppara$) in terms of Hermite-Laguerre gyromoments.
By expressing the canonical polynomial basis $\{x^n\}$ for $n=0,1,2$ into Hermite and Laguerre polynomials, we deduce for the gyrofluid density
$n_a = N_a^{00}$, for the parallel pressure $\Ppara=\sqrt{2}N_a^{20}+N_a^{00}$, for the perpendicular pressure $\Pperpa=N_a^{00}-N_a^{01}$, and for the components of the energy-weighted pressure tensor $\Rparpara=\sqrt{3/2}N_a^{40}/2 - 3\sqrt{2} N_a^{20}/4 - 3 N_a^{00}/8$, $\Rxa = -N_a^{21}/\sqrt{2}$, and $\Rperpperpa = N_a^{02}$. 
We notice that the other gyrofluid moments, i.e. the fluid velocity and heat flux, vanish due to the symmetry rising from the $\kpar=0$ assumption in the Z-pinch geometry.
In terms of gyrofluid moments, Eqs. (\ref{eq:momenthierarchy_start}-\ref{eq:momenthierarchy_21_end}) write

\begin{align}
    \Dt{n_a} &+ \frac{\tau_a}{z_a}ik_y(\Ppara + \Pperpa)     
    % \nonumber \\ &\qquad \qquad
    = \squareparenthesis{\roundparenthesis{\kappa_N - 2}\kernel_0 +(1-\kappa_T)\kernel_1}i k_y \phi \label{eq:GMGF_dens}\\
    \Dt{\Ppara} &+ \frac{\tau_a}{z_a}ik_y\roundparenthesis{-4n_a+7\Ppara+\Pperpa + 2\Rparpara + \Rxa}
    \nonumber \\ &\quad
    = \squareparenthesis{\roundparenthesis{\kappa_N+\kappa_T-4}\kernel_0+\roundparenthesis{1-\kappa_T}\kernel_1}i k_y \phi\label{eq:GMGF_ppar}\\
    \Dt{\Pperpa} &+ \frac{\tau_a}{z_a}ik_y\roundparenthesis{-3n_a + \Ppara + 5\Pperpa + \Rxa + \Rperpperpa}
    \nonumber \\ &\quad
    = \squareparenthesis{(\kappa_N+\kappa_T-3)\kernel_0+(5-\kappa_N-3\kappa_T)\kernel_1+(2\kappa_N-2)\kernel_2}i k_y \phi\label{eq:GMGF_pperp}.
    % %
    % \Dt{\Rparpara} &+ \frac{\tau_a}{z_a}ik_y\roundparenthesis{-\frac{3}{2}\Ppara -\frac{3}{2}\Pperpa + 4\Rparpara -3\Rxa}
    % \nonumber \\ &\quad
    % = \frac{3}{2}\squareparenthesis{\roundparenthesis{6 - \kappa_N - 2\kappa_T}\kernel_0+\roundparenthesis{\kappa_T-1}\kernel_1}i k_y \phi\\
    % %
    % \Dt{\Rxa} &+ \frac{\tau_a}{z_a}ik_y\roundparenthesis{-3n_a+\Ppara+2\Pperpa + 8\Rxa}
    % = \roundparenthesis{\kappa_T-2}\kernel_1 i k_y \phi\\
    % %
    % \Dt{\Rperpperpa} &+ \frac{\tau_a}{z_a}ik_y\roundparenthesis{-4n_a + 4\Pperpa + 6\Rperpperpa} \label{eq:GMGF_Rperperpa}
    % = [4\kernel_1 - 12 \kernel_2] i k_y \phi.
\end{align}
The linear terms in Eqs. (\ref{eq:GMGF_dens}-\ref{eq:GMGF_pperp}) are equivalent to the gyrofluid equations presented in \cite{Brizard1992NonlinearPlasmas}, by replacing $L_\perp=L_B$ and considering electrostatic fluctuations in the Z-pinch geometry (i.e. $\eta_B=1$, $\epsilon_\beta=0$, $V_\parallel=0$, $\partial_\parallel = 0$ and $\omega_\grad = \omega_\kappa$, adapting \cite{Brizard1992NonlinearPlasmas} notations).
Expressing the equilibrium FLR differential operator as $\Delta_\perp = -b_a$, we deduce for the first kernels $\kernel_0=e^{\Delta_\perp}$, $\kernel_1=-\Delta_\perp e^{\Delta_\perp}$, $\kernel_2=\Delta_\perp^2 e^{\Delta_\perp}/2$. 
Thus, by setting higher-order kernels to zero, we retrieve the FLR correction terms present in the gyrofluid model.
% It is worth noting that expanding the gyromoment equations up to the energy pressure tensor moments, i.e. $\{N_a^{40},N_a^{21},N_a^{02}\}$ or $\{\Rparpara,\Rxa,\Rperpperpa\}$, modifies slightly the linear coefficients for density and parallel pressure dependencies in Eq. \eqref{eq:GMGF_ppar}.
% One can obtain the correct coefficients if one does not express the energy pressure tensor moments.\\
Finally, the total derivatives $\Dt{.}$ yield the nonlinear gyrofluid terms
\begin{align}
    \Dt{n_a} &= \ddt n_a + \{\kernel_0\phi,n_a\} + \{\kernel_1\phi,n_a-\Pperpa\} + \frac{1}{2}\{\kernel_2\phi,\Rperpperpa\} \label{eq:GMGF_NL_dens}\\
    \Dt{\Ppara}&= \ddt\Ppara + \{\kernel_0\phi,\Ppara\} + \{\kernel_1\phi,n_a-\Pperpa\}
        \nonumber \\&\qquad
        -\{\kernel_1\phi, \Rxa\} + \frac{1}{2}\{\kernel_2\phi,\Rperpperpa+\sqrt{2} N_a^{22}\}\label{eq:GMGF_NL_ppar}\\
    \Dt{\Pperpa}&= \ddt\Pperpa + \{\kernel_0\phi,\Pperpa\} +\{\kernel_1\phi,2n_a-3\Pperpa\} + \{\kernel_2\phi,2\Pperpa-2n_a\}
        \nonumber \\&\qquad
        -\{\kernel_1\phi,\Rperpperpa\} + \{\kernel_2\phi,5/2\Rperpperpa+3 N_a^{03}\}\label{eq:GMGF_NL_pperp}.
    % %
    % S^{\Rparpara}=\ \{\kernel_0\phi,\Rparpara\}&-\frac{3}{2}\{\kernel_1\phi,n_a-\Pperpa - 2\Rxa + a_3N_a^{41}\}
    % \nonumber \\
    % &+\{\kernel_2\phi,-3/4 \Rperpperpa + a_4 N_a^{22} + a_5 N_a^{42}\}\\
    % %
    % S^{\Rxa}= \{\kernel_0\phi,\Rxa\} &+\{\kernel_1\phi,n_a-\Ppara - 2\Rxa + a_6 N_a^{22}\}
    % \nonumber \\
    % &+\{\kernel_2\phi,2\Rxa - a_7 N_a^{22}+ a_8 N_a^{23} \}\\
    % %
    % S^{\Rperpperpa}= \{\kernel_0\phi,\Rperpperpa\}&+4\{\kernel_1\phi,n_a - \Pperpa - \Rperpperpa + a_9 N_a^{02}\}+\{\kernel_2\phi,2\Rxa\}
    % \nonumber \label{eq:GMGF_NL_pperp}\\
    % &+\{\kernel_2\phi,-6n_a+8\Pperpa + 10\Rperpperpa - a_{10} N_a^{03} + a_{11} N_a^{04}\}
\end{align}
In \cite{Brizard1992NonlinearPlasmas}, only density and temperature fluctuations contribute to the nonlinear terms.
This is equivalent to set $\Rparpara=\Rxa=\Rperpperpa=N_a^{22}=N_a^{03}=0$ in Eqs. (\ref{eq:GMGF_NL_dens}-\ref{eq:GMGF_NL_pperp}).
Thus, in this limit, our gyromoment hierarchy, Eqs. (\ref{eq:GMGF_dens}-\ref{eq:GMGF_pperp}) is equivalent to the $\delta f$ gyrofluid framework for the description of the ion dynamics.
The electrons are modeled adopting a drift-kinetic limit in the gyrofluid equations, which can be easily obtained from our gyromoment framework by setting $\kernel_n = \delta_{n0}$.

\subsection{Relation with Hasegawa-Mima model}
The \cite{Hasegawa1978Pseudo-three-dimensionalPlasma} and \cite{Hasegawa1983} models consider cold ions, adiabatic electrons, $z_a=\tau_a=1$ and the long wavelength limit $\kperp \ll 1$.
By applying these assumptions, the gyromoment hierarchy, Eq. \eqref{eq:GK_moment_hierarchy}, reduces to a single moment model
\begin{equation}
    \ddt N_i^{00}+\{N_i^{00},\phi\}+ 2 i k_y n_i^{00} -ik_y \kappa_N\phi = 0
    \label{eq:one_moment}
\end{equation}
where the $\kperp\ll 1$ limit allows us to approximate $\kernel_n = \delta_{n0}$.
Introducing the ion perturbed density, $n=N_i^{00}$, and defining $\varphi=-\phi$, Eq. \eqref{eq:one_moment} writes in real space
\begin{equation}
    \ddt n + \{\varphi,n\} + \kappa_N\ddy \varphi - \kappa_B\ddy(n-\varphi) = 0,
    \label{eq:density_HM}
\end{equation}
where we set $\kappa_B = 2R/L_\perp$.
Using the modified adiabatic electron response $n_e = \phi - \langle \phi \rangle_z = 0$ in a two-dimensional Z-pinch, the Poisson equation yields $n_i=\grad^2\varphi$.
This leads to an extended Hasegawa-Mima equation for the vorticity, $\gradperp^2 \varphi$, containing magnetic curvature and gradient effects
\begin{gather}
    \ddt (\gradperp^2 \varphi) + \{\varphi,\gradperp^2 \varphi\} + \kappa_N\ddy \gradperp^2 \varphi - \kappa_B\ddy(\gradperp^2 \varphi-\varphi) = 0   .  \label{eq:HM}
    % \zeta = \grad^2 \varphi\nonumber.
\end{gather}
Eq. \eqref{eq:HM} correspond to the model deduced by \cite{Dewhurst2009TheTurbulence} when imposing $\kpar=0$ or neglecting the resistive coupling between $\varphi$, $n$ and the parallel current.
}

\bibliographystyle{jpp}
\bibliography{references}

\begin{thebibliography}{66}
\expandafter\ifx\csname natexlab\endcsname\relax\def\natexlab#1{#1}\fi
\def\au#1{#1} \def\ed#1{#1} \def\yr#1{#1}\def\at#1{#1}\def\jt#1{\textit{#1}}
  \def\bt#1{#1}\def\bvol#1{\textbf{#1}} \def\vol#1{#1} \def\pg#1{#1}
  \def\publ#1{#1}\def\arxiv#1{#1}\def\org#1{#1}\def\st#1{\textit{#1}}

\bibitem[Abel {\em et~al.\/}(2008)Abel, Barnes, Cowley, Dorland \&
  Schekochihin]{Abel2008LinearizedTheory}
{\sc \au{Abel, I.~G.}, \au{Barnes, M.}, \au{Cowley, S.~C.}, \au{Dorland, W.} \&
  \au{Schekochihin, A.~A.}} \yr{2008}  \at{{Linearized model Fokker-Planck
  collision operators for gyrokinetic simulations. I. Theory}}.  \jt{Physics of
  Plasmas}  \bvol{15}~(12).

\bibitem[Armstrong(1967)]{Armstrong1967NumericalEquation}
{\sc \au{Armstrong, Thomas~P.}} \yr{1967}  \at{{Numerical studies of the
  nonlinear Vlasov equation}}.  \jt{Physics of Fluids}  \bvol{10}~(6),
  \pg{1269--1280}.

\bibitem[Barnes {\em et~al.\/}(2009)Barnes, Abel, Dorland, Ernst, Hammett,
  Ricci, Rogers, Schekochihin \& Tatsuno]{Barnes2009}
{\sc \au{Barnes, M.}, \au{Abel, I.~G.}, \au{Dorland, W.}, \au{Ernst, D.~R.},
  \au{Hammett, G.~W.}, \au{Ricci, P.}, \au{Rogers, B.~N.}, \au{Schekochihin,
  A.~A.} \& \au{Tatsuno, T.}} \yr{2009}  \at{{Linearized model Fokker-Planck
  collision operators for gyrokinetic simulations. II. Numerical implementation
  and tests}}.  \jt{Physics of Plasmas}  \bvol{16}~(7),  \pg{072107}.

\bibitem[Beer {\em et~al.\/}(1995)Beer, Cowley \& Hammett]{Beer1995}
{\sc \au{Beer, M.~A.}, \au{Cowley, S.~C.} \& \au{Hammett, G.~W.}} \yr{1995}
  \at{{Field-aligned coordinates for nonlinear simulations of tokamak
  turbulence}}.  \jt{Physics of Plasmas}  \bvol{2}~(7),  \pg{2687--2700}.

\bibitem[Belli \& Candy(2012)]{Belli2012FullSimulations}
{\sc \au{Belli, E.~A.} \& \au{Candy, J.}} \yr{2012}  \at{{Full linearized
  Fokker-Planck collisions in neoclassical transport simulations}}.  \jt{Plasma
  Physics and Controlled Fusion}  \bvol{54}~(1).

\bibitem[Brizard(1992)]{Brizard1992NonlinearPlasmas}
{\sc \au{Brizard, Alain}} \yr{1992}  \at{{Nonlinear gyrofluid description of
  turbulent magnetized plasmas}}.  \jt{Physics of Fluids B}  \bvol{4}~(5),
  \pg{1213--1228}.

\bibitem[Brizard \& Hahm(2007)]{Brizard2007}
{\sc \au{Brizard, A.~J.} \& \au{Hahm, T.~S.}} \yr{2007}  \at{{Foundations of
  nonlinear gyrokinetic theory}}.  \jt{Reviews of Modern Physics}
  \bvol{79}~(2),  \pg{421--468}.

\bibitem[Brunner {\em et~al.\/}(2000)Brunner, Valeo \&
  Krommes]{Brunner2000LinearTransport}
{\sc \au{Brunner, S.}, \au{Valeo, E.} \& \au{Krommes, J.~A.}} \yr{2000}
  \at{{Linear delta-f simulations of nonlocal electron heat transport}}.
  \jt{Physics of Plasmas}  \bvol{7}~(7),  \pg{2810--2823}.

\bibitem[Catto(1978)]{Catto1978LinearizedGyro-kinetics}
{\sc \au{Catto, P.~J.}} \yr{1978}  \at{{Linearized gyro-kinetics}}.  \jt{Plasma
  Physics}  \bvol{20}~(7),  \pg{719--722}.

\bibitem[Dewhurst {\em et~al.\/}(2009)Dewhurst, Hnat \&
  Dendy]{Dewhurst2009TheTurbulence}
{\sc \au{Dewhurst, J.~M.}, \au{Hnat, B.} \& \au{Dendy, R.~O.}} \yr{2009}
  \at{{The effects of nonuniform magnetic field strength on density flux and
  test particle transport in drift wave turbulence}}.  \jt{Physics of Plasmas}
  \bvol{16}~(7),  \pg{072306}.

\bibitem[Diamond {\em et~al.\/}(2005)Diamond, Itoh, Itoh \& Hahm]{Diamond2005}
{\sc \au{Diamond, P.~H.}, \au{Itoh, S.~I.}, \au{Itoh, K.} \& \au{Hahm, T.~S.}}
  \yr{2005}  \at{{Zonal flows in plasma - A review}}.  \jt{Plasma Physics and
  Controlled Fusion}  \bvol{47}~(5).

\bibitem[Dimits {\em et~al.\/}(2000)Dimits, Bateman, Beer, Cohen, Dorland,
  Hammett, Kim, Kinsey, Kotschenreuther, Kritz, Lao, Mandrekas, Nevins, Parker,
  Redd, Shumaker, Sydora \& Weiland]{Dimits2000ComparisonsSimulations}
{\sc \au{Dimits, A.~M.}, \au{Bateman, G.}, \au{Beer, M.~A.}, \au{Cohen, B.~I.},
  \au{Dorland, W.}, \au{Hammett, G.~W.}, \au{Kim, C.}, \au{Kinsey, J.~E.},
  \au{Kotschenreuther, M.}, \au{Kritz, A.~H.}, \au{Lao, L.~L.}, \au{Mandrekas,
  J.}, \au{Nevins, W.~M.}, \au{Parker, S.~E.}, \au{Redd, A.~J.}, \au{Shumaker,
  D.~E.}, \au{Sydora, R.} \& \au{Weiland, J.}} \yr{2000}  \at{{Comparisons and
  physics basis of tokamak transport models and turbulence simulations}}.
  \jt{Physics of Plasmas}  \bvol{7}~(3),  \pg{969--983}.

\bibitem[Dougherty(1964)]{Dougherty1964}
{\sc \au{Dougherty, J.~P.}} \yr{1964}  \at{{Model Fokker-Planck Equation for a
  Plasma and Its Solution}}.  \jt{Physics of Fluids}  \bvol{7}~(11),
  \pg{1788}.

\bibitem[Frei {\em et~al.\/}(2021)Frei, Ball, Hoffmann, Jorge, Ricci \&
  Stenger]{Frei2021b}
{\sc \au{Frei, B.~J.}, \au{Ball, J.}, \au{Hoffmann, A.C.D.}, \au{Jorge, R.},
  \au{Ricci, P.} \& \au{Stenger, L.}} \yr{2021}  \at{{Development of advanced
  linearized gyrokinetic collision operators using a moment approach}}.
  \jt{Journal of Plasma Physics}  \bvol{87}~(5),  \pg{905870501}.

\bibitem[Frei {\em et~al.\/}(2022{\natexlab{{\em a\/}}})Frei, Hoffmann \&
  Ricci]{Frei2022LocalMode}
{\sc \au{Frei, B.~J.}, \au{Hoffmann, A.C.D.} \& \au{Ricci, P.}}
  \yr{2022{\natexlab{{\em a\/}}}}  \at{{Local gyrokinetic collisional theory of
  the ion-temperature gradient mode}}.  \jt{Journal of Plasma Physics}
  \bvol{88}~(3),  \pg{905880304}.

\bibitem[Frei {\em et~al.\/}(2022{\natexlab{{\em b\/}}})Frei, Hoffmann, Ricci,
  Brunner \& Tecchiolli]{Frei2022Moment-BasedModel}
{\sc \au{Frei, B.~J.}, \au{Hoffmann, A. C.~D.}, \au{Ricci, P.}, \au{Brunner,
  S.} \& \au{Tecchiolli, Z.}} \yr{2022{\natexlab{{\em b\/}}}}
  \at{{Moment-Based Approach to the Flux-Tube linear Gyrokinetic Model}}.
  \jt{Under consideration for publication in J. Plasma Phys.
  (arXiv:2210.05799v1)} .

\bibitem[Frei {\em et~al.\/}(2020)Frei, Jorge \& Ricci]{Frei2020}
{\sc \au{Frei, B.~J.}, \au{Jorge, R.} \& \au{Ricci, P.}} \yr{2020}  \at{{A
  gyrokinetic model for the plasma periphery of tokamak devices}}.  \jt{Journal
  of Plasma Physics}  \bvol{86}~(2),  \pg{905860205}.

\bibitem[Frieman \& Chen(1982)]{Frieman1982NonlinearEquilibria}
{\sc \au{Frieman, E.~A.} \& \au{Chen, Liu}} \yr{1982}  \at{{Nonlinear
  gyrokinetic equations for low-frequency electromagnetic waves in general
  plasma equilibria}}.  \jt{Physics of Fluids}  \bvol{25}~(3),  \pg{502--508}.

\bibitem[Frigo \& Johnson(2005)]{FFTW05}
{\sc \au{Frigo, M.} \& \au{Johnson, S.~G.}} \yr{2005}  \at{{The Design and
  Implementation of FFTW3}}.  \jt{Proceedings of the IEEE}  \bvol{93}~(2),
  \pg{216--231}.

\bibitem[Fujisawa {\em et~al.\/}(2004)Fujisawa, Itoh, Iguchi, Matsuoka,
  Okamura, Shimizu, Minami, Yoshimura, Nagaoka, Takahashi, Kojima, Nakano,
  Ohsima, Nishimura, Isobe, Suzuki, Akiyama, Ida, Toi, Itoh \&
  Diamond]{Fujisawa2004}
{\sc \au{Fujisawa, A.}, \au{Itoh, K.}, \au{Iguchi, H.}, \au{Matsuoka, K.},
  \au{Okamura, S.}, \au{Shimizu, A.}, \au{Minami, T.}, \au{Yoshimura, Y.},
  \au{Nagaoka, K.}, \au{Takahashi, C.}, \au{Kojima, M.}, \au{Nakano, H.},
  \au{Ohsima, S.}, \au{Nishimura, S.}, \au{Isobe, M.}, \au{Suzuki, C.},
  \au{Akiyama, T.}, \au{Ida, K.}, \au{Toi, K.}, \au{Itoh, S.~I.} \&
  \au{Diamond, P.~H.}} \yr{2004}  \at{{Identification of zonal flows in a
  toroidal plasma}}.  \jt{Physical Review Letters}  \bvol{93}~(16),  \pg{1--4}.

\bibitem[Gibelli \& Shizgal(2006)]{Gibelli2006SpectralTerm}
{\sc \au{Gibelli, Livio} \& \au{Shizgal, Bernie~D.}} \yr{2006}  \at{{Spectral
  convergence of the Hermite basis function solution of the Vlasov equation:
  The free-streaming term}}.  \jt{Journal of Computational Physics}
  \bvol{219}~(2),  \pg{477--488}.

\bibitem[Goldston \& Rutherford(1995)]{Goldston1995IntroductionPhysics}
{\sc \au{Goldston, R.} \& \au{Rutherford, P.}} \yr{1995} {\em {Introduction to
  Plasma Physics}\/}.  \publ{Institute of Physics Publishing}.

\bibitem[Gradshteyn \& Ryzhik(2014)]{Gradshteyn2014TableProducts}
{\sc \au{Gradshteyn, I.S.} \& \au{Ryzhik, I.M.}} \yr{2014} {\em {Table of
  Integrals, Series, and Products}\/}.  \publ{Academic Press Inc.}

\bibitem[Grant \& Feix(1967)]{Grant1967Fourier-HermiteLimit}
{\sc \au{Grant, Frederick~C.} \& \au{Feix, Marc~R.}} \yr{1967}
  \at{{Fourier-Hermite solutions of the Vlasov equations in the linearized
  limit}}.  \jt{Physics of Fluids}  \bvol{10}~(4),  \pg{696--702}.

\bibitem[Hallenbert \& Plunk(2021)]{Hallenbert2021PredictingLimit}
{\sc \au{Hallenbert, A.} \& \au{Plunk, G.~G.}} \yr{2021}  \at{{Predicting the
  Dimits shift through reduced mode tertiary instability analysis in a strongly
  driven gyrokinetic fluid limit}}.  \jt{Journal of Plasma Physics}
  \bvol{87}~(5),  \pg{905870508}.

\bibitem[Hallenbert \& Plunk(2022)]{Hallenbert2022}
{\sc \au{Hallenbert, A.} \& \au{Plunk, G.~G.}} \yr{2022}  \at{{Predicting the
  Z-pinch Dimits shift through gyrokinetic tertiary instability analysis of the
  entropy mode}}.  \jt{Journal of Plasma Physics}  \bvol{88}~(4),
  \pg{905880402}.

\bibitem[Hammett {\em et~al.\/}(1992)Hammett, Dorland \&
  Perkins]{Hammett1992FluidDynamics}
{\sc \au{Hammett, G.~W.}, \au{Dorland, W.} \& \au{Perkins, F.~W.}} \yr{1992}
  \at{{Fluid models of phase mixing, landau damping, and nonlinear gyrokinetic
  dynamics}}.  \jt{Physics of Fluids B}  \bvol{4}~(7),  \pg{2052--2061}.

\bibitem[Hasegawa \& Mima(1978)]{Hasegawa1978Pseudo-three-dimensionalPlasma}
{\sc \au{Hasegawa, Akira} \& \au{Mima, Kunioki}} \yr{1978}
  \at{{Pseudo-three-dimensional turbulence in magnetized nonuniform plasma}}.
  \jt{Physics of Fluids}  \bvol{21}~(1),  \pg{87--92}.

\bibitem[Hasegawa \& Wakatani(1983)]{Hasegawa1983}
{\sc \au{Hasegawa, Akira} \& \au{Wakatani, Masahiro}} \yr{1983}  \at{{Plasma
  Edge Turbulence}}.  \jt{Physical Review Letters}  \bvol{50}~(9),
  \pg{682--686}.

\bibitem[Hazeltine \& Meiss(2003)]{Hazeltine2003PlasmaConfinement}
{\sc \au{Hazeltine, R.D.} \& \au{Meiss, J.D.}} \yr{2003} {\em {Plasma
  Confinement}\/}, corrected republ. edn.  \publ{Mineola, New York: Dover
  Publ.}

\bibitem[Helander \& Sigmar(2002)]{Helander2002CollisionalPlasmas}
{\sc \au{Helander, P.} \& \au{Sigmar, D.J.}} \yr{2002} {\em {Collisional
  Transport in Magnetized Plasmas}\/}.  \publ{Cambridge University Press}.

\bibitem[Held {\em et~al.\/}(2020)Held, Wiesenberger \&
  Kendl]{Held2020Pade-basedModels}
{\sc \au{Held, M.}, \au{Wiesenberger, M.} \& \au{Kendl, A.}} \yr{2020}
  \at{{Pad{\'{e}}-based arbitrary wavelength polarization closures for full-F
  gyro-kinetic and -fluid models}}.  \jt{Nuclear Fusion}  \bvol{60},
  \pg{066014}.

\bibitem[Hoffmann \& Frei(2020)]{gyacomo}
{\sc \au{Hoffmann, A.C.D.} \& \au{Frei, B.J.}} \yr{2020} {The GYACOMO code: a
  nonlinear gyrokinetic advanced collision moment solver,
  gitlab.epfl.ch/ahoffman/gyacomo}.

\bibitem[Ivanov {\em et~al.\/}(2022)Ivanov, Schekochihin \&
  Dorland]{Ivanov2022}
{\sc \au{Ivanov, P.~G.}, \au{Schekochihin, A.~A.} \& \au{Dorland, W.}}
  \yr{2022}  \at{{Dimits transition in three-dimensional
  ion-temperature-gradient turbulence}}.  \jt{Accepted in Journal of Plasma
  Physics} .

\bibitem[Ivanov {\em et~al.\/}(2020)Ivanov, Schekochihin, Dorland, Field \&
  Parra]{Ivanov2020ZonallyTurbulence}
{\sc \au{Ivanov, Plamen~G.}, \au{Schekochihin, A.~A.}, \au{Dorland, W.},
  \au{Field, A.~R.} \& \au{Parra, F.~I.}} \yr{2020}  \at{{Zonally dominated
  dynamics and Dimits threshold in curvature-driven ITG turbulence}}.
  \jt{Journal of Plasma Physics}  \bvol{86}~(5),  \pg{855860502}.

\bibitem[Jenko {\em et~al.\/}(2000)Jenko, Dorland \&
  Kotschenreuther]{Jenko2000}
{\sc \au{Jenko, F.}, \au{Dorland, W.} \& \au{Kotschenreuther, M.}} \yr{2000}
  \at{{Electron Temperature Gradient Driven Turbulence}}.  \jt{Physics of
  Plasmas}  \bvol{7}~(5),  \pg{1904--1910}.

\bibitem[Jorge {\em et~al.\/}(2019)Jorge, Frei \& Ricci]{Jorge2019b}
{\sc \au{Jorge, R.}, \au{Frei, B.~J.} \& \au{Ricci, P.}} \yr{2019}
  \at{{Nonlinear gyrokinetic Coulomb collision operator}}.  \jt{Journal of
  Plasma Physics}  \bvol{85}~(6),  \pg{1--31}.

\bibitem[Jorge {\em et~al.\/}(2017{\natexlab{{\em a\/}}})Jorge, Ricci \&
  Loureiro]{Jorge2017ACollisionality}
{\sc \au{Jorge, R.}, \au{Ricci, P.} \& \au{Loureiro, N.~F.}}
  \yr{2017{\natexlab{{\em a\/}}}}  \at{{A drift-kinetic analytical model for
  scrape-off layer plasma dynamics at arbitrary collisionality}}.  \jt{Journal
  of Plasma Physics}  \bvol{83}~(6).

\bibitem[Jorge {\em et~al.\/}(2017{\natexlab{{\em b\/}}})Jorge, Ricci \&
  Loureiro]{Jorge2017}
{\sc \au{Jorge, R.}, \au{Ricci, P.} \& \au{Loureiro, N.~F.}}
  \yr{2017{\natexlab{{\em b\/}}}}  \at{{A drift-kinetic analytical model for
  scrape-off layer plasma dynamics at arbitrary collisionality}}.  \jt{Journal
  of Plasma Physics}  \bvol{83}~(6),  \pg{905830606}.

\bibitem[Joyce {\em et~al.\/}(1971)Joyce, Knorr \&
  Meier]{Joyce1971NumericalEquation}
{\sc \au{Joyce, Glenn}, \au{Knorr, Georg} \& \au{Meier, Homer~K.}} \yr{1971}
  \at{{Numerical integration methods of the Vlasov equation}}.  \jt{Journal of
  Computational Physics}  \bvol{8}~(1),  \pg{53--63}.

\bibitem[Kobayashi \& G{\"{u}}rcan(2015)]{Kobayashi2015a}
{\sc \au{Kobayashi, S.} \& \au{G{\"{u}}rcan, Ö.D.}} \yr{2015}
  \at{{Gyrokinetic turbulence cascade via predator-prey interactions between
  different scales}}.  \jt{Physics of Plasmas}  \bvol{22}~(5),  \pg{050702}.

\bibitem[Kobayashi {\em et~al.\/}(2015)Kobayashi, G{\"{u}}rcan \&
  Diamond]{Kobayashi2015}
{\sc \au{Kobayashi, S.}, \au{G{\"{u}}rcan, Ö.~D.} \& \au{Diamond, P.~H.}}
  \yr{2015}  \at{{Direct identification of predator-prey dynamics in
  gyrokinetic simulations}}.  \jt{Physics of Plasmas}  \bvol{22}~(9),
  \pg{090702}.

\bibitem[Kobayashi \& Rogers(2012)]{Kobayashi2012}
{\sc \au{Kobayashi, S.} \& \au{Rogers, B.~N.}} \yr{2012}  \at{{The quench rule,
  Dimits shift, and eigenmode localization by small-scale zonal flows}}.
  \jt{Physics of Plasmas}  \bvol{19}~(1),  \pg{012315}.

\bibitem[Lenard \& Bernstein(1958)]{Lenard1958}
{\sc \au{Lenard, A.} \& \au{Bernstein, I.}} \yr{1958}  \at{{Plasma Oscillations
  with Diffusion in Velocity Space}}.  \jt{Physical Review E}  \bvol{11}~(12),
  \pg{1456--1459}.

\bibitem[Lin {\em et~al.\/}(1999)Lin, Hahm, Lee, Tang \&
  Diamond]{Lin1999EffectsTransport}
{\sc \au{Lin, Z.}, \au{Hahm, T.~S.}, \au{Lee, W.~W.}, \au{Tang, W.~M.} \&
  \au{Diamond, P.~H.}} \yr{1999}  \at{{Effects of collisional zonal flow
  damping on turbulent transport}}.  \jt{Physical Review Letters}
  \bvol{83}~(18),  \pg{3645--3648}.

\bibitem[Loureiro {\em et~al.\/}(2016)Loureiro, Dorland, Fazendeiro, Kanekar,
  Mallet, Vilelas \& Zocco]{Loureiro2016Viriato:Dynamics}
{\sc \au{Loureiro, N.~F.}, \au{Dorland, W.}, \au{Fazendeiro, L.}, \au{Kanekar,
  A.}, \au{Mallet, A.}, \au{Vilelas, M.~S.} \& \au{Zocco, A.}} \yr{2016}
  \at{{Viriato: A Fourier-Hermite spectral code for strongly magnetized
  fluid-kinetic plasma dynamics}}.  \jt{Computer Physics Communications}
  \bvol{206},  \pg{45--63}.

\bibitem[Madsen(2013)]{Madsen2013Full-FModel}
{\sc \au{Madsen, Jens}} \yr{2013}  \at{{Full-F gyrofluid model}}.  \jt{Physics
  of Plasmas}  \bvol{20}~(7).

\bibitem[Manas {\em et~al.\/}(2017)Manas, Hornsby, Angioni, Camenen \&
  Peeters]{Manas2017ImpactSimulations}
{\sc \au{Manas, P.}, \au{Hornsby, W.~A.}, \au{Angioni, C.}, \au{Camenen, Y.} \&
  \au{Peeters, A.~G.}} \yr{2017}  \at{{Impact of the neoclassical distribution
  function on turbulent impurity and momentum fluxes: Fluid model and
  gyrokinetic simulations}}.  \jt{Plasma Physics and Controlled Fusion}
  \bvol{59}~(3).

\bibitem[Mandell {\em et~al.\/}(2022)Mandell, Dorland, Abel, Gaur, Kim, Martin
  \& Qian]{Mandell2022GX:Design}
{\sc \au{Mandell, N.~R.}, \au{Dorland, W.}, \au{Abel, I.}, \au{Gaur, R.},
  \au{Kim, P.}, \au{Martin, M.} \& \au{Qian, T.}} \yr{2022}  \at{{GX: a
  GPU-native gyrokinetic turbulence code for tokamak and stellarator design}}.
  \jt{Under consideration for publication in J. Plasma Phys.
  (arXiv:2209.06731v3)} .

\bibitem[Mandell {\em et~al.\/}(2018{\natexlab{{\em a\/}}})Mandell, Dorland \&
  Landreman]{Mandell2018}
{\sc \au{Mandell, N.~R.}, \au{Dorland, W.} \& \au{Landreman, M.}}
  \yr{2018{\natexlab{{\em a\/}}}}  \at{{Laguerre – Hermite pseudo-spectral
  velocity formulation of gyrokinetics}}.  \jt{Journal of Plasma Physics}
  \bvol{84}~(1),  \pg{905840108}.

\bibitem[Mandell {\em et~al.\/}(2018{\natexlab{{\em b\/}}})Mandell, Dorland \&
  Landreman]{Mandell2018TheGx.readthedocs.io}
{\sc \au{Mandell, N.~R.}, \au{Dorland, W.} \& \au{Landreman, M}}
  \yr{2018{\natexlab{{\em b\/}}}} {The GX code: gx.readthedocs.io}.

\bibitem[Orszag(1971)]{Orszag1971}
{\sc \au{Orszag, S.~A.}} \yr{1971}  \at{{On the Elimination of Aliasing in
  Finite-Difference Schemes by Filtering High-Wavenumber Components}}.
  \jt{Journal of the Atmospheric Sciences}  \bvol{28}~(6),  \pg{1074--1074}.

\bibitem[Parker \& Dellar(2015)]{Parker2015Fourier-HermiteLimit}
{\sc \au{Parker, Joseph~T.} \& \au{Dellar, Paul~J.}} \yr{2015}
  \at{{Fourier-Hermite spectral representation for the Vlasov-Poisson system in
  the weakly collisional limit}}.  \jt{Journal of Plasma Physics}
  \bvol{81}~(2).

\bibitem[Pueschel {\em et~al.\/}(2010)Pueschel, Dannert \&
  Jenko]{Pueschel2010OnMicroturbulence}
{\sc \au{Pueschel, M.~J.}, \au{Dannert, T.} \& \au{Jenko, F.}} \yr{2010}
  \at{{On the role of numerical dissipation in gyrokinetic Vlasov simulations
  of plasma microturbulence}}.  \jt{Computer Physics Communications}
  \bvol{181}~(8),  \pg{1428--1437}.

\bibitem[Qi {\em et~al.\/}(2020)Qi, Majda \& Cerfon]{Qi2020}
{\sc \au{Qi, D.}, \au{Majda, A.~J.} \& \au{Cerfon, A.~J.}} \yr{2020}
  \at{{Dimits shift, avalanche-like bursts, and solitary propagating structures
  in the two-field flux-balanced Hasegawa-Wakatani model for plasma edge
  turbulence}}.  \jt{Physics of Plasmas}  \bvol{27}~(10),  \pg{102304}.

\bibitem[Ricci {\em et~al.\/}(2006{\natexlab{{\em a\/}}})Ricci, Rogers \&
  Dorland]{Ricci2006a}
{\sc \au{Ricci, P.}, \au{Rogers, B.~N.} \& \au{Dorland, W.}}
  \yr{2006{\natexlab{{\em a\/}}}}  \at{{Small-scale turbulence in a
  closed-field-line geometry}}.  \jt{Physical Review Letters}  \bvol{97}~(24),
  \pg{8--11}.

\bibitem[Ricci {\em et~al.\/}(2010)Ricci, Rogers \& Dorland]{Ricci2010}
{\sc \au{Ricci, P.}, \au{Rogers, B.~N.} \& \au{Dorland, W.}} \yr{2010}
  \at{{Collisional damping of zonal flows due to finite Larmor radius
  effects}}.  \jt{Physics of Plasmas}  \bvol{17}~(7),  \pg{1--9}.

\bibitem[Ricci {\em et~al.\/}(2006{\natexlab{{\em b\/}}})Ricci, Rogers, Dorland
  \& Barnes]{Ricci2006}
{\sc \au{Ricci, P.}, \au{Rogers, B.~N.}, \au{Dorland, W.} \& \au{Barnes, M.}}
  \yr{2006{\natexlab{{\em b\/}}}}  \at{{Gyrokinetic linear theory of the
  entropy mode in a Z pinch}}.  \jt{Physics of Plasmas}  \bvol{13}~(6),
  \pg{062102}.

\bibitem[Rogers \& Dorland(2005)]{Rogers2005}
{\sc \au{Rogers, B.~N.} \& \au{Dorland, W.}} \yr{2005}
  \at{{Noncurvature-driven modes in a transport barrier}}.  \jt{Physics of
  Plasmas}  \bvol{12}~(6),  \pg{062511}.

\bibitem[Rosenbluth \& Longmire(1957)]{Rosenbluth1957}
{\sc \au{Rosenbluth, M.~N.} \& \au{Longmire, C.~L.}} \yr{1957}  \at{{Stability
  of plasmas confined by magnetic fields}}.  \jt{Annals of Physics}
  \bvol{1}~(2),  \pg{120--140}.

\bibitem[Scott(2005)]{scott2005}
{\sc \au{Scott, Bruce~D.}} \yr{2005}  \at{{Drift wave versus interchange
  turbulence in tokamak geometry: Linear versus nonlinear mode structure}}.
  \jt{Physics of Plasmas}  \bvol{12}~(6),  \pg{1--23}.

\bibitem[Smith(1991)]{Smith1991AlgorithmArithmetic}
{\sc \au{Smith, D.~M.}} \yr{1991}  \at{{Algorithm 693: A FORTRAN package for
  floating-point multiple-precision arithmetic}}.  \jt{ACM Transactions on
  Mathematical Software (TOMS)}  \bvol{17}~(2),  \pg{273--283}.

\bibitem[Snyder \& Hammett(2001)]{Snyder2001AMicroturbulence}
{\sc \au{Snyder, P.~B.} \& \au{Hammett, G.~W.}} \yr{2001}  \at{{A Landau fluid
  model for electromagnetic plasma microturbulence}}.  \jt{Physics of Plasmas}
  \bvol{8}~(7),  \pg{3199--3216}.

\bibitem[Strintzi {\em et~al.\/}(2005)Strintzi, Scott \&
  Brizard]{Strintzi2005NonlocalModel}
{\sc \au{Strintzi, D.}, \au{Scott, B.~D.} \& \au{Brizard, A.~J.}} \yr{2005}
  \at{{Nonlocal nonlinear electrostatic gyrofluid equations: A four-moment
  model}}.  \jt{Physics of Plasmas}  \bvol{12},  \pg{052517}.

\bibitem[Sugama {\em et~al.\/}(2009)Sugama, Watanabe \&
  Nunami]{Sugama2009LinearizedEquations}
{\sc \au{Sugama, H.}, \au{Watanabe, T.~H.} \& \au{Nunami, M.}} \yr{2009}
  \at{{Linearized model collision operators for multiple ion species plasmas
  and gyrokinetic entropy balance equations}}.  \jt{Physics of Plasmas}
  \bvol{16}~(11),  \pg{112503}.

\bibitem[Zocco \& Schekochihin(2011)]{Zocco2011}
{\sc \au{Zocco, A.} \& \au{Schekochihin, A.~A.}} \yr{2011}  \at{{Reduced
  fluid-kinetic equations for low-frequency dynamics, magnetic reconnection,
  and electron heating in low-beta plasmas}}.  \jt{Physics of Plasmas}
  \bvol{18}~(10),  \pg{102309}.

\end{thebibliography}
\end{document}